\documentclass{pasa}%
\usepackage{natbib}
\usepackage{url}
\usepackage{graphicx}
\usepackage{amsmath} 
\usepackage{graphicx}
\usepackage{amssymb}
\usepackage{amstext}
\usepackage{floatpag}
\usepackage{float}
\def \prd {PRD}
\def \apj {ApJ}
\def \apjs {ApJS}
\def \apjl {ApJL}
\def \mnras {MNRAS}
\def \aap {A\&A}

\def \apss {APSS}

\def \nat {Nature}
\def \physrep {Phys. Rep.}
\def \nar {New. A. Rev.}
\def \apss {Ap\&SS}

\def \pasj {PASJ}
\def \pasa {PASA}

\def \pasp {PASP}

\usepackage{amsmath} 

\newcommand{\degree}{$^{\circ}$}
\newcommand{\msol}{$M_{\odot}$}
\newcommand{\msoleq}{M_{\odot}}

\title[Australia's role in multi-messenger astronomy]{Hunting Gravitational Waves with Multi-Messenger Counterparts: Australia's Role}
\author[E. J. Howell et al.]{E. J. Howell$^{1}$\thanks{E-mail:eric.howell@uwa.edu.au}, A. Rowlinson$^{2,8}$, D. M. Coward$^{1}$, P. D. Lasky$^{3,4}$, D. L. Kaplan$^5$, E. Thrane$^{3,4}$, G. Rowell$^6$, \mbox{D.~K. Galloway$^{3,4}$,} Fang Yuan$^{7,8}$, R. Dodson$^9$, T. Murphy$^{8,10}$, G. C. Hill$^6$, I. Andreoni$^{11}$, L. Spitler$^{12,13}$ and A. Horton$^{12,13}$\vspace{2mm}\\
\affil{$^1$School of Physics, University of Western Australia, Crawley WA 6009, Australia}
\affil{$^2$CSIRO Astronomy and Space Science, Sydney, Australia}
\affil{$^3$Monash Centre for Astrophysics, Monash University, VIC 3800, Australia}
\affil{$^4$School of Physics \& Astronomy, Monash University, VIC 3800, Australia}
\affil{$^5$Department of Physics, University of Wisconsin–-Milwaukee, Milwaukee, WI 53201, USA}
\affil{$^6$Department of Physics, School of Physical Sciences, University of Adelaide, Adelaide, SA 5005, Australia}
\affil{$^{7}$Research School of Astronomy and Astrophysics,Australian National University,Canberra, ACT 2611, Australia}
\affil{$^{8}$ARC Centre of Excellence for All-sky Astrophysics (CAASTRO)}
\affil{$^9$International Centre for Radio Astronomy Research, M468, The University of Western Australia, Crawley, WA 6009, Australia}
\affil{$^{10}$Sydney Institute for Astronomy (SIfA), School of Physics, The University of Sydney, NSW 2006, Australia}
\affil{$^{11}$Centre for Astrophysics and Supercomputing, Swinburne University of Technology, Hawthorn VIC 3122, Australia}
\affil{$^{12}$Department of Physics \& Astronomy, Macquarie University, Sydney, NSW 2109, Australia}
\affil{$^{13}$Australian Astronomical Observatories, PO Box 915 North Ryde NSW 1670, Australia}
}

\jid{PASA}

\begin{document}%

\begin{abstract}
The first observations by a worldwide network of advanced interferometric gravitational wave detectors offer a unique opportunity for the  astronomical  community. At design sensitivity, these facilities will be able to detect coalescing binary neutron stars to distances approaching \mbox{400 Mpc}, and neutron star-black hole systems to \mbox{1 Gpc}. Both of these sources are associated with gamma ray bursts which are known to emit across the entire electromagnetic spectrum. Gravitational wave detections provide the opportunity for  ``multi-messenger'' observations, combining gravitational wave with electromagnetic, cosmic ray or neutrino observations. This review provides an overview of how Australian astronomical facilities and collaborations with the gravitational wave community can contribute to this new era of discovery, via contemporaneous follow-up observations from the radio to the optical and high energy. We discuss some of the frontier discoveries that will be made possible when this new window to the Universe is opened.
\end{abstract}

\maketitle%

%
\section{Astronomy in the Gravitational Wave Era}
\label{sec:intro}

The detection of gravitational waves (GWs) will rank as one of the major scientific achievements of this century. Their detection will open up a new observational window to the Universe, revealing dynamic sources of strong field relativistic gravity previously inaccessible through conventional astronomical instruments. Our understanding of space-time and matter under the most extreme conditions will be transformed.

Although there has been no direct detection of GWs to date, indirect evidence for their existence comes from high precision, Nobel-prize winning measurements of the pulsar PSR 1913+16 and its companion neutron star  \citep[NS; ][]{Hulse_Taylor_75,Weisberg_Taylor_84}. The GW emission that drives the system's orbital decay is in agreement with the predictions of general relativity to better than 1\% \citep{Hartle03}.

When such binary neutron star (BNS) systems eventually coalesce, they are predicted to emit copious amounts of GWs \citep{thorn}. These sources will be prime targets for the new generation of GW detectors, led by Advanced LIGO \citep[aLIGO;][]{aLIGO_2015} which is set to begin observing during the second half of 2015 and Advanced Virgo a year later \citep[][]{Acernese_2015}. At final sensitivity, these advanced detectors are expected to detect BNS mergers at a rate within the range 0.4--400 yr$^{-1}$ \citep{Abadie2010CQGra}. Compact Binary Coalescences (CBCs) consisting of at least one black hole (BH) are also targets for GW detectors; although there is compelling evidence for their existence \citep{Barnard_IC10X-1_2008,Prestwich_BH_IC10-X1_07}, the event rates of these sources for aLIGO detection is not well known.

One realisation in the last decade is that coalescing systems of NS/NS or NS/BH events could be the progenitors of short-hard gamma ray bursts (SGRBs); transient events routinely observed throughout the electromagnetic (EM) spectrum \citep{Paczynski1986ApJ,Eichler1989Natur,Narayan_1992ApJ,Rezzolla_2011ApJ,gehrels_2005,Berger2005Natur, Bloom2006ApJ}. There exist other types of EM, neutrino and cosmic ray emissions that may also be associated with GW events. These include long-duration gamma ray bursts \citep[LGRBs;][]{Kobayashi2003ApJ}, short gamma ray repeaters \citep{Abbott2008PhRvL}, supernovae \citep{Fryer2002ApJ,Ott_ccSNreview_09}, fast radio bursts \citep{Zhang2014ApJ} as well as others.

History has already shown that multi-wavelength astronomy can play an important role in unveiling new phenomena. In the last decade, X-ray, optical and radio follow-ups have all transformed and revealed new processes in our understanding of gamma ray bursts (GRBs); combining EM observations with those in the GW domain will too provide new insight into the internal engines and mechanisms at play in a multitude of different sources. A new generation of sensitive, wide-field telescopes, advancements in time domain astronomy and upgrades to neutrino and cosmic ray detectors can provide a coordinated network for discovery. The possible simultaneous detection of photons, neutrinos or high energy particles with GWs would be a landmark moment for astrophysics, initiating a new era of \emph{multi-messenger}\footnote{The term multi-messenger stems from the various type of \emph{messengers} that can arrive from different astrophysical events; other than EM photons, these can include particles such as neutrinos, cosmic rays or indeed GWs.} astronomy, for the first time including GW.

Maximising the potential offered by GW observations involves the development of a worldwide, multi-messenger network.  Australian facilities are ideally placed to foster scientific exchanges in this new era and agreements have already been established. To conduct EM follow-up of GW triggers, memorandums of understanding (MoUs) have been signed between the LIGO/Virgo GW collaboration and a number of facilities either based in Australia or with strong Australian involvement; these include: The Anglo-Australian Telescope, the Australian Square Kilometer Array Pathfinder \citep[ASKAP;][]{Murphy2013PASAa}, the Cherenkov Telescope Array \citep[CTA;][]{Acharya_2013APh}, The High Energy Stereoscopic System \citep[H.E.S.S;][]{Lennarz2013}, IceCube \citep{IC2006APh}, The Murchison Widefield Array \citep[MWA;][]{Tingay2013PASA}, and the SkyMapper \citep{Keller2007PASA}, The GW Optical Transient Observer (GOTO\footnote{\url{http://goto-observatory.org/}}) and Zadko \citep{Coward2010PASA} optical telescopes.

In this paper, we focus on the most probable multi-messenger observations from the advanced detector era; those associated with GRBs. Whilst doing so, we consider the contribution that the Australian facilities can make to the worldwide multi-messenger effort.

The structure of this paper is as follows: Section 2 describes GW astronomy. Sections 3 and 4 introduce SGRBs and LGRBs and describe how co-ordinated GW and multiwavelength observations of these events can provide breakthrough science. Section 5 acts as a primer for those unfamiliar with the concepts and terminologies of detection and data analysis often used in the GW domain; this section is not designed to be exhaustive but to present some of the most important concepts in GW detection and data analysis. Section 6 discusses the expected rates and detection ranges for GW sources. The next two sections describe two of the strategies that form the basis for coordinated GW and EM observations in the GW era. Section 7 discusses EM triggered GW searches; these could likely yield the first coincident GW-EM event through archival GW data. Section 8 discusses the  EM follow-up of GW Triggers;
this strategy is highly challenging due to the large positional uncertainties of GW observations but the potential rewards for success are without doubt highly significant. Section 10 discusses the Australian facilities involved in the co-ordinated science programs with aLIGO/AdV and we highlight the areas in which they could contribute in this new frontier.  Finally, in section 11 we discuss the role neutrino follow-up plays in GW detection.

\section{Gravitational waves: a new type of Astronomy}
Gravitational waves are produced by regions of space-time that are distorted by high velocity bulk motions of matter. The timescale of the motions determine the frequency of the GW emission; ground based detectors will target systems with masses in the range 1--10$^{3}$ \msol, which emit in the 1~Hz--10~kHz band. This frequency range, covering the audio band, has motivated the characterisation of interferometric GW astronomy as  ``listening to the Universe''.

Instruments capable of achieving detections will begin observations in the second half of 2015. Advanced LIGO, a pair of US based interferometric detectors at Hanford and Livingston \citep[USA][]{aLIGO_2015} will have its first observational science run (O1) in late-2015; a year later it will be joined by the Italian Advanced Virgo \citep[AdV;][]{Acernese_2015,virgo2012JInst} for a second observing run (O2).
The ``advanced'' network of interferometric GW detectors will eventually have 10 times the sensitivity of the first generation instruments. The increased sensitivity translates into a factor $10^{3}$ increase in observed volume, making detections expected rather than plausible.

Additional instruments are expected to eventually join the network. KAGRA, a Japanese detector, is envisioned to begin operation in 2018-19 \citep{kagra_2012} and LIGO-India is expected to be operational from 2020, reaching a design sensitivity at the same level as aLIGO by around 2022~\citep{LSC_Prospects_aLIGO_2013}.

The GW observations made by these instruments will differ from most conventional EM observations in several ways:
\begin{enumerate}

\item[$\bullet$] GWs are not scattered or obscured by intervening material like dust so provide a window into the densest regions of the Universe.

\item[$\bullet$] As GW detectors observe an amplitude rather than a flux, the measure of detectability follows an inverse relationship with distance rather than the conventional inverse square law. Therefore, number counts of a homogeneous distribution of standard-candle sources increases with distance, $d$, as, $d^{3}$, rather than, $d^{3/2}$;

\item[$\bullet$] As GWs couple weakly to the detectors, even very local astronomical sources of GWs have to be highly energetic emitters of gravitational radiation;

\item[$\bullet$] GW detectors are nearly omnidirectional, with a nearly $4\pi$ steradian sensitivity to astrophysical events with a greater than average response over more than 40\% of the sky.

\end{enumerate}
The first point implies that GW observations can allow us to view astrophysical phenomena inaccessible by other means. The gravitational window can therefore enable frontier explorations in the low to intermediate redshift universe ($z \lesssim 0.4$) of sources that are electromagnetically invisible for much, or all, of their lives. The second point means that a factor 2 improvement in the sensitivity of a GW detector results in a factor 8 increase in the volume of the Universe being probed. The third point emphasises a detection bias for detecting the most highly energetic astrophysical events. The typical fluxes of GW sources are of order $10^{20}$ Jy, far greater than equivalent fluxes typically observed in the radio domain ($\mu$Jy -- Jy). The final point means that GW detectors are naturally survey instruments over a wide band of frequencies (10--5000 Hz).

There are a number of types of EM counterparts that may be associated with GW emissions \citep{Branchesi2012JPhCS,Mandel_2012IAUS}. As some of these counterparts are quite speculative, this paper focuses on GW signals associated with GRBs. Other sources of simultaneous EM and GW emission include supernovae as well as multiple emission mechanisms from NSs; for a review of the latter, see the accompanying article in this series \citep{lasky15pasa}.

In the next few sections we provide a summary of both SGRBs and LGRBs and the type of GW/EM associations that could be targeted in the GW era. Some of these predictions are based on solid foundations whilst some are more speculative. In considering the latter, we note that when a new window of observation has been opened in the past, the discoveries that transform our understanding of the Universe have often been the least expected.

\section{Multi-Messenger Astronomy with Short Gamma Ray Bursts}
\begin{figure*}
\includegraphics[scale = 0.13,origin=rl]{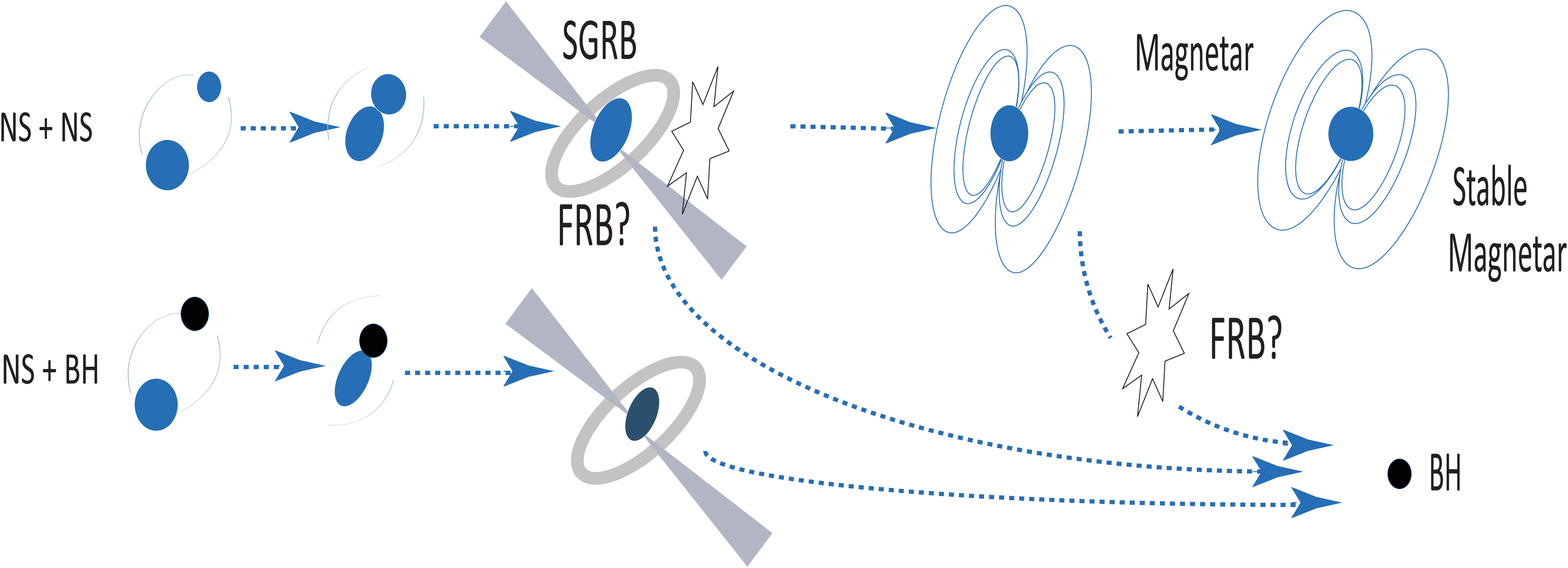}\\
  \caption{A cartoon illustrating some of the possible scenarios for coalescing systems of NSs and BHs. Short duration gamma ray bursts (SGRBs) have been linked with the merger of compact objects \citep{Berger2005Natur, Bloom2006ApJ} and could be accompanied by a fast radio burst \citep[FRB; ][]{Thornton2013Sci,Lorimer2013MNRAS,Totani2013PASJ,Palaniswamy2014ApJ,Zhang2014ApJ}. If a stable magnetar is formed, the long lived X-ray plateaus observed in many SGRBs could indicate a constant energy injection \citep{Corsi2009ApJ,Rowlinson2010MNRAS,Rowlinson2013MNRAS,Zhang2013ApJ,Gao2013,Fan2013PhRvD}; the possible collapse of a merger product to a BH could also result in an FRB \citep{Falcke_Rezzolla2014AA,Zhang2014ApJ}. Figure adapted from Chu et al. 2015.}
  \label{fig_mm_scenarios}
\end{figure*}

Gravitational waves from the merger of coalescing binary systems of NS/NS and NS/BHs\footnote{If the BH mass is greater than 10 times the NS mass, the NS will be swallowed without leaving any residual disc \citep{miller_05,Pannarale2014ApJ,Maselli_PhysRevD_2014}.} are confidently predicted to have observable EM counterparts. This expectation is a result of the connection between these events and SGRBs \citep[e.g.][]{Eichler1989Natur,gehrels_2005,Tanvir2013Natur,Berger2013ApJ}. The evidence stems from a number of different channels. Firstly, the dynamic timescales of discs predicted from the merger of CBCs are consistent with the durations of SGRBs. Secondly, EM follow-ups of SGRBs have never provided an associated supernova. Thirdly, SGRB afterglows have been localised to galaxies harbouring older stellar populations with offsets of order tens of kpc from their galactic centers; this is consistent with post-natal kick velocities of 100s of km s$^{-1}$, and also with the fainter and shorter lived afterglows expected from an ambient interstellar medium at a large offset. Finally, as is discussed in \S \ref{kilonova}, the discovery of a faint EM transient called a \emph{kilonova} has provided the strongest observational evidence to date of the SGRB/CBC association.

Conclusive proof of the CBC/SGRB association will be provided through GW observations. Coincident EM and GW observations of SGRBs could also provide a fascinating insight to the dominant mechanisms at the heart of GRBs. Low-latency GW pipelines could enable multi-wavelength follow-up measurements of the prompt emission, constraining both the underlying central engines and the emission mechanisms at work \citep{Elliott2014A&A}. Later-time multi-wavelength follow-ups can provide insight through extensive coverage of the SGRB afterglow.

A number of EM counterparts have been predicted to accompany the inspiral and merger of NS/NS and NS/BH systems. In Figure \ref{fig_mm_scenarios} we show the likely outcomes of these mergers and in the following sections we will briefly review the most likely EM counterparts that could accompany CBCs.

\subsection{Prompt emission}
\label{section_prompt}
During the final stages of the merger of a compact binary, the system is expected to launch a highly relativistic jet that interacts with itself and the surrounding medium \citep[the fireball model for GRBs; e.g.][]{Piran_1999}. Collisions of material moving at different velocities within the jet will lead to internal shocks, giving short-lived bursts of gamma-rays that we detect as the SGRB prompt emission. As the accretion timescale is expected to be $<$2 seconds \citep{Metzger_2008}, the GRBs associated with compact binary mergers are typically shorter in duration than those associated with core-collapse supernovae \citep[explaining the observed distribution of GRBs;][]{Kouveliotou_1993}. However, the division between these two populations is not easily identifiable from the prompt gamma-ray emission alone \citep[e.g.][]{Bromberg_2013,Qin_2013}.

A number of {\it Fermi}-LAT GRBs have shown $>$GeV emission (even at redshifts as distant as $z\approx4$). Two bursts were observed with gamma-ray photons reaching energies up to 94\,GeV (GRB 130427A) and 62\,GeV (GRB 131231A) ---  this supports the suggestion that the photon energies may extend higher than previously assumed \citep{Bouvier_2011,Inoue_2013APh}. Significantly, these discoveries have not been limited to LGRBs, with SGRBs also showing high energy photons and GeV emission  often continuing for 10's of seconds beyond the initial burst. The fact that {\it Fermi}-LAT discovered a photon of energy 31\,GeV during the prompt phase of GRB 090510 \citep{Ackermann_2010ApJ} is promising for co-ordinated observations between GW detectors and ground-based Cherenkov telescope arrays \citep{Bartos_2014MNRAS} operating at $>$ 10 GeV. Additionally, SGRBs with time-extended emission have recently been cited as promising targets for Cherenkov telescope arrays \citep{Veres_2014ApJ}.

One exciting possibility is the observation of prompt optical flashes. So far, these emissions have only been observed in LGRBs \citep{Racusin2008Natur,Akerlof1999Natur,Vestrand2014Sci}. An early optical emission correlated with the prompt gamma-rays could indicate a common origin related to the internal shocks \citep{Vestrand2005Natur}.

A number of studies have suggested that compact binary mergers could generate prompt coherent radio emission \citep[e.g.,][]{Totani2013PASJ}.
Such mechanisms include excitation of the plasma surrounding a compact binary merger by GWs \citep{Moortgat_2005}, from a dynamically-generated magnetic field after the merger \citep{Pshirkov2010},
or from the onset of the collision of the forward shock with the surrounding medium \citep{Usov_2000,Sagiv_2002}. However, the detectability of emission from these processes will be dependent upon the scattering by the surrounding environment \citep{Macquart_2007}. Nonetheless, these studies suggest compact binary mergers are an interesting contender for the progenitors of Fast Radio Bursts \citep[FRBs;][]{Lorimer_2007,Thornton2013Sci},
which are currently unknown.

\subsection{Energy injection at late times}
\label{section_magnetar}
Plateaus and flares in X-ray light curves following GRBs are signatures of ongoing energy injection.  This could be caused by late-time accretion onto a central black hole \citep[unlikely in the compact binary scenario; see discussion in][]{Rowlinson2013MNRAS}, or from ongoing energy injection from the spindown of a newly born neutron star.  Indeed, recent studies \citep[e.g.,][]{Giacomazzo2013ApJ,Zhang2013ApJ,Lasky2014PhRvD} have shown that the merger of two NSs could result in a supramassive NS; a star with a mass greater than the non-rotating maximum mass but supported from further collapse through rotation \citep{Cook1994ApJ}.

Around 60\% of X-ray afterglow light curves of SGRBs observed by the \emph{Swift} satellite\footnote{http://heasarc.gsfc.nasa.gov/docs/swift/swiftsc.html} \citep{Gehrels_Swift_2004} have shown plateaus lasting 100--10000~s after the burst; these have been attributed to electromagnetic spin-down emissions from protomagnetars \citep{Rowlinson2010MNRAS,Rowlinson2013MNRAS} formed via the merger of two neutron stars \citep{Dai1998,Zhang2001ApJ}. Observations of the plateau phase can also be used to constrain the NS equation of state, with GW observations of the inspiral phase significantly aiding this endeavour \citep{Lasky2014PhRvD}

If the post-merger remnant is an NS, early optical afterglow as bright as 17th magnitude in R band (assuming a distance of $\sim300$ Mpc; see \S \ref{section_CBC_Ranges}) could be produced from dissipation of a wide-beamed protomagnetar wind \cite{Zhang2013ApJ}.  This magnetar wind could launch ejecta at relativistic speeds which would interact with the surrounding medium and produce a bright broadband afterglow from synchrotron radiation \citep{Gao2013}.

Gravitational wave emission may also accompany an afterglow plateau if a millisecond magnetar is born from the collision.  Multiple mechanisms for generating such GWs exist in nascent neutron stars, including secular bar modes \citep[e.g., ][]{LaiShapiro_1995,shibataBarmode04,Corsi:2009}, $r$-modes \citep{Andersson1998ApJ,Andersson2001IJMPD}, and magnetic-field induced stellar deformations \citep{Cutler2002PhRvD,Haskell2008MNRAS,DallOsso2015ApJ}.  Such emission could be observable by aLIGO out to $\lesssim100$ Mpc \citep{Corsi:2009,Fan2013PhRvD,DallOsso2015ApJ}.  In fact, the X-ray light curve itself can be used to constrain the total GW emission from these systems \citep{Lasky2015inprep}.

Some plateaus following SGRBs exhibit an extremely steep decay phase, commonly interpreted as the collapse of the nascent neutron star to a black hole \citep[][]{Troja_2007,Lyons_2010,Rowlinson2010MNRAS, Rowlinson2013MNRAS}.  Such collapse could potentially produce an FRB when the magnetic field lines snap as they cross the BH horizon \citep{Falcke_Rezzolla2014AA,Zhang2014ApJ}, which is expected to occur $\lesssim5\times10^{4}$ s after the merger \citep{Ravi2014MNRAS}. A low latency GW trigger could enable prompt follow-ups to test this connection \citep{Chu_inprep}.

\subsection{Afterglow}
As the relativistic jet propagates, it collides with the medium surrounding the progenitor resulting in a forward shock travelling into the surrounding medium, and a reverse shock propagating back up the jet \citep[e.g.][]{Sari_1999,Rees_1992}. These shock fronts produce multi-wavelength synchrotron emission, initially peaking in the X-ray and moving through the different wavelengths to radio as it fades. The typical afterglow of GRBs is attributed to the forward shock emission and the brightness of this afterglow is dependent upon a number of parameters, including the density of the surrounding medium. Therefore, in a low density environment, the forward shock component is expected to be relatively faint.

The multi-wavelength afterglows of SGRBs have been observed and are typically fainter than those of LGRBs \citep[][]{Berger_2007,Gehrels_2008,Nysewander_2009,Kann_2011}. This is consistent with SGRBs being less energetic than LGRBs and with their locations in lower density environments. The reverse shock has also been observed for SGRB 051221A \citep[e.g.][]{Soderberg_2006}.

\subsection{Kilonova}
\label{kilonova}
A `kilonova' is been predicted to form after the merger of two NSs. This faint optical transient is powered by the radioactive decay of the ejected neutron rich matter \citep{Li1998ApJ,Rosswog2005ApJ,Metzger2010MNRAS} and could reach around 21–-23 mag in the optical and 21–-24 mag in the NIR for a source at 200 Mpc \citep{Tanaka2013ApJ}. Recent optical and near-infrared follow-up observations of GRB 130603B have provided the most conclusive evidence to date of this scenario, reinforcing the theory that compact object mergers are the progenitors of SGRBs \citep{Tanvir2013Natur,Berger2013ApJ}. These observations have added significantly to other observational evidence in support of this scenario \citep{Berger2005Natur,Bloom2006ApJ,Berger_2009ApJ}. Coincident EM and GW observations could confirm that SGRBs are indeed the result of coalescing compact binaries.

An additional prompt EM emission related to the kilonova mechanism has also recently been suggested by \citet{Metzger_2014}. This has been inspired by studies that suggest a small fraction of the ejected neutron rich matter can expand rapidly enough to avoid r-process capture \citep{Bauswein2013ApJ}. The suggestion is that $\beta$-decay from free neutrons in the outermost layers of this ejecta could power optical emission on a timescale of hours after the merger, peaking at around magnitude 22 in the U-band for a source. For a source at 200 Mpc this signal would peak at around magnitude 22 in the U-band and would act as a precursor to a kilonova.

\section{Long Gamma Ray Bursts as multi-messenger targets for GWs}
\label{section_LGRBs}
LGRBs are amongst the most luminous transient events in the Universe in terms of EM radiation per unit solid angle. These beamed emissions have been observed to last up to $10^{4}$ s \citep{Gendre2013ApJ,Greiner_2015} and can radiate a total energy equivalent to that of the Sun in its entire 10 Gyr lifetime. The extreme luminosities allow LGRBs to be seen out to cosmological volumes, making them a probe of the high redshift universe ($z > 5$).

The favoured scenario for these bursts is described by the collapsar model \citep{Woosley_99} in which the inner part of a Wolf-Rayet star progenitor collapses to form a rapidly rotating black hole. High angular momentum enables the infalling matter to form an accretion disk, which in turn provides the energy reservoir to power an ultra-relativistic jet that blasts its way through the stellar envelope. The observed radiation is explained through synchrotron and/or inverse Compton emission from the accelerated electrons in internal and external shocks. Some authors have  suggested instead that the central engines may consist of magnetars \citep{Usov:1992zd,DuncanThomson_92,Bucciantini_2009}. There is observational evidence to support this scenario for at least a proportion of LGRBs \citep{Metzger2011MNRAS}.

The connection between LGRBs with the collapse of massive stars \citep{WB_06,Hjorth_LGRB_SN,Stanek_LGRB_SN} has been supported by afterglow observations in or near dense regions of active star-formation; predominantly dwarf starburst field galaxies \citep{Fruchter_grb_gal}. As mentioned earlier, their denser environments, as well as their higher emission energies, mean that the multiwavelength afterglows of LGRBs are typically brighter than those that occur from SGRBs \citep[][]{Nysewander_2009,Kann_2011}.

In terms of GW emissions from these events, a number of LGRBs have been associated with core-collapse supernova \citep{Hjorth_LGRB_SN,Campana2006Natur}. Modelling the GW emission from these supernovae is very complex, requiring a combination of general relativistic hydrodynamics, magnetic fields, rotation, neutrino transport and nuclear physics \citep[][]{Ott_ccSNreview_09}. Simulations have so far provided a picture of a very complex and chaotic behavior that includes shock formation and turbulence that create highly complex waveforms with multiple sharp bursts over ms durations. However, most scenarios suggest an event may have to be within tens of kpc for detection. As most LGRBs occur at cosmological distances, the vast majority of their GW signals will be out of reach for advanced detectors.

The requirement for rapid rotation to produce the disc in a GRB \citep{WoosleyJankaNature_05} allows for alternative emission mechanisms that could produce detectable GWs out to 10s of Mpc \citep{Fryer2011}. Fragmentation instabilities could be produced in the core or in the disc \citep{Fryer2002ApJ,Kobayashi2003ApJ}. Rapid rotation could also give rise to
rotational instabilities in the protoneutron star remnant \citep{Dimmelmeier2008PhRvD,Corsi:2009,Piro_Ott_2011,Piro_Thrane_2012}

A number of studies have suggested there exists a sub-population of LGRBs known as low-luminosity GRBs (\emph{ll}GRBs). These events have isotropic equivalent gamma-ray luminosities 2-3 orders of magnitude below classical LGRBs \citep{coward_LLGRB_05,murase_06,GuettaDellaValle_2007,Imerito_08,howell_2013} and have only been detected at low-$z$ due to their lower energy emissions (the closest was GRB\,980425 at z = 0.0085 or 36Mpc). As such their rates have been predicted to be 2--3 orders of magnitude greater than LGRBs.

Observations have confirmed that both LGRBs and \emph{ll}GRBs produce supernovae, suggesting that the \emph{ll}GRBs may just be lower-energy events from the tail of the distribution. This has been a long going debate and attempts to address it have used statistical arguments \citep{sodoburg_06_LLGRBRate_06,GuettaDellaValle_2007}, fits to the peak flux distribution \citep{Pian_LLGRBs_06}, and simulation \citep{coward_LLGRB_05,Virgilii_LLGRBs_08}. The suggestion that \emph{ll}GRBs could be just normal LGRBs viewed off-axis was discounted based on statistical arguments, as it would produce a far higher local rate density than expected from LGRBs and would require narrower opening angles for LGRBs than determined from the breaks in afterglow lightcurves \citep{Daigne_2007}.

Recently, an analysis of \emph{ll}GRB 060218 has suggested that the main difference in the two bursts arises from an extended low-mass envelope in \emph{ll}GRBs \citep{Nakar2015ApJ}. The existence of such an envelope can smother the jet and drive a mildly relativistic shock resulting in a much lower luminosity than that produced by an ultra-relativistic jet that is able to penetrate through the bare progenitor star. Interestingly, the statistical arguments suggesting separate populations put forward by \citet{howell_2013} can also support these two different scenarios. It is therefore possible that GW emission mechanisms could be driven by the same type of engine for both these classes.

\section{GW sensitivity and Networks}
\subsection{Instrument sensitivity}
The output from a single GW detector consists of a time series data stream, $s(t)$, composed of the detector response to a GW signal, $h(t)$, and the detector noise $n(t)$:
\begin{equation}
\noindent
s(t) = h(t) + n(t)
       \hspace{0.5mm}.\label{eq_h_antenna_response}
\end{equation}
In general, $h(t)$ will be a linear combination of the two orthogonal transverse polarizations, $h_{+,\hspace{0.5mm}\times}$, weighted by the dimensionless detector antenna pattern functions for the two polarizations $F_{{+},{\times}}$:
\begin{equation}
\hspace{-0.5mm}
h(t) = F_{+}(t,\hspace{0.5mm}\theta,\hspace{0.5mm}\phi,\hspace{0.5mm}\psi )\hspace{0.5mm}h_{+}(t) + F_{\times}(t,\hspace{0.5mm}\theta, \hspace{0.5mm}\phi,\hspace{0.5mm}\psi )\hspace{0.5mm}h_{\times}(t)
       \,,\label{eq_h_antenna_response}
\end{equation}
\noindent which describe the detector sensitivity to radiation of different polarizations, incident from different directions \citep{shultz_antenna_patt_fun_87,tinto_antenna_patt_fun_87,jaranowski_anttenapatfuns_98}. The angles, $\theta$ and $\phi$, represent the direction to the source and $\psi$ is the polarization angle of the wave.

A GW detector can follow the phase of a GW signal, so the time series is generally represented in the frequency domain by the strain amplitude spectral density, $\tilde{h}(f)$. This quantity is defined through the power spectral density $S_{s}(f)=\tilde{s}^{*}(f)\tilde{s}(f)$, with
$\tilde{s}(f)$ the Fourier transform of the time series. Similarly, one can define a signal power spectral density, $S_{h}(f)$, and a noise power spectral density, $S_{n}(f)$. The strain amplitude spectral density is given by:
\begin{equation}\label{eq_h_f}
    \tilde{h}(f) = \sqrt{S_{s}(f)}\hspace{1mm}\,,
\end{equation}
\noindent with dimensions of $\mathrm{Hz}^{-1/2}$ \citep{thorn}. This quantity is often used in plots to display the sensitivity of GW interferometers.

\subsection{GW detector networks}
\label{GW detector networks}
A single GW detector can not determine the polarization state or source direction of a transient signal\footnote{For a continuous wave source, directionality can be obtained from Doppler modulations of the signal due to the movement of the detector relative to the source.}. To obtain source localization, a widely separated network of GW detectors is essential. Such a network may employ techniques such as \emph{coincidence analysis}, in which individual events from different detectors are correlated in time \citep{Arnaud02}, or \emph{coherent analysis}, in which synchronized detector outputs are merged before searching for a common pattern \citep{Finn02}. By effectively resolving the different times of arrival of GW events between members of a network, coherent network analysis enables a detector array to become an all-sky monitor with good angular resolution over all source directions.

Achieving good directional sensitivity is of paramount importance for GW/EM associations. For the sources considered in this review, directional sensitivity is determined through triangulation of arrival times\footnote{Typically the angular resolution of a GW network is inversely proportional to the separation of the detectors in the network.}. To maximize the time delays, and hence improve directionality, it is advantageous that a network be as geographically widely separated as possible \citep{Saty04} and as such, a number of detectors are planned to join the aLIGO/AdV network throughout the next decade.

The Japanese observatory KAGRA\footnote{This was previously known as LCGT. KAGRA derives the "KA" from its location at the Kamioka mine and "GRA" from gravity.}, should begin operations by around 2018-19 \citep{KAGRA}; at design sensitivity this detector could improve the directional precision of a aLIGO/AdV network by a factor of 1.5-2 and the detection rate by a similar factor \citep{Fairhurst2011CQGra,Chu_inprep}. LIGO-India operating at aLIGO sensitivity will be added to the aLIGO/AdV network by 2022 --- by then BNSs will be detectable out to $\sim$200 Mpc and up to 400 events are possible per year \citep{Abadie2010CQGra}. An Indian detector will improve the angular resolution sufficiently to increase the percentage of GW sources detected within 5 deg$^{2}$ from 3-7\% to 17\% \citep{LSC_Prospects_aLIGO_2013}.

It has long been recognized that a GW detector in Australia would add the longest baseline to the proposed advanced detector network \citep[e.g.,][]{cavalier06,blair08, wen10}. For example, adding an Australian detector to an aLIGO/AdV three detector network can reduce the error in solid angle to tens of arc-minutes for high signal-to-noise ratio (SNR) signals \citep{advGWDet}, dramatically improving the ability to localise GW sources for multi-wavelength follow-up observations. This scenario could be realised when third generation observatories such as the `Einstein gravitational wave Telescope' (ET)\footnote{http://www.et-gw.eu/} become a reality in the next decade \citep{Hild_2008,Hild_2010,Hild_2011}. The optimal site for a detector in the southern hemisphere been shown to be Western Australia \citep{Schutz2011CQGra}, the current home to an 80-m baseline prototype GW detector\footnote{AIGO---\url{http://www.aigo.org.au/aigores.php}}.

\subsection{The GW false alarm rate}
\label{section_far}
The false alarm rate (FAR) is the rate that false positives appear above a given SNR threshold, and is dependent on the number of glitches (non-stationary transients) in the GW data stream. It is a critical measure as it determines whether a candidate should be considered for follow-up. For well modeled sources, the background of false alarms is at a level close to that of Gaussian noise. For un-modeled sources --- typically short duration transients --- the data quality has a greater effect on detection confidence. One therefore sets the threshold high enough so that noise generated false alarms are negligible. Given that the probability, $P(h)\rm{d}h$, of observing an event with an amplitude in the range $h$ to $h + \mathrm{d}h$ is given by a Gaussian distribution of standard deviation $\sigma$, the probability of obtaining a FAR greater than a given threshold, $\rho$, is:
\begin{equation}\label{eq_far}
  P(h |h > z) = \frac{1}{\sqrt{2 \pi} \sigma} \int^{\infty}_{\rho} \rm{exp}\left(  \frac{-h^{2}}{2 \sigma^2} \right ) \rm{d}h \,.
\end{equation}

To be 99\% confident that a GW has been detected, one can set an SNR $\sim 8 $ which is equivalent to a FAR of 1 in 100 years of observation \mbox{($3 \times 10^{-10} $\,Hz )}. To see this one can approximate number of noise instances during that period. If the detector output sampling rate is 1 kHz and the output is processed through $\sim 10^{3}$ filters, in 100 years we get $P(h |h > z) = (3 \times 10^{15})^{-1} $ , yielding $\rho \sim 8 $ which is our required SNR \citep[see][for a detailed discussion of this argument] {SathyaprakashSchutz_LIVREV}. For a network of three equivalent detectors combined SNR, $\rho_{c}$ is given as:
\begin{equation}\label{eq_combined_snr}
    \rho_{c} = \sqrt{\sum_{i}{\rho^{2}_{i}}},
\end{equation}
\noindent where $\rho_{i}$ represents the SNR in the ith detector \citep{cutler94}. This shows that for a network of 3 equivalent detectors, to dismiss false alarms at a level $3 \times 10^{-10} $\,Hz requires $\rho_{c} \sim 12$.
\begin{figure}
  \includegraphics[scale = 0.55]{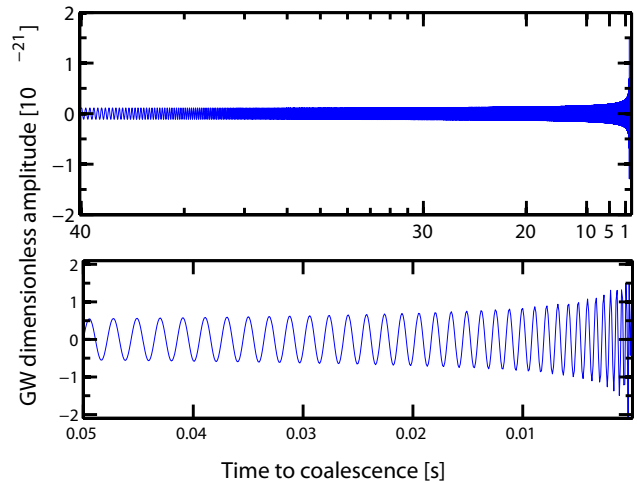}\\
  \caption{\textbf{Top:} The  predicted chirp waveform of a coalescing compact binary system 40s before merger. As the signal increases in both amplitude and frequency towards merger, it will sweep across the sensitive bandwidth of advanced GW interferometric detectors. After the merger, the signal will show a ring down phase (not shown in this plot) which will take the form of a increasingly damped sinusoid. \textbf{Bottom:} The final 50ms before merger.}
  \label{fig_chirp}
\end{figure}

\section{Gravitational waves from Inspiralling Compact Objects}
\subsection{Detection of inspiralling compact objects}
\label{section_cbc}
The expected GW signals from CBCs takes on the well modelled chirp form shown in Figure \ref{fig_chirp}. The figure shows how the signal increases in both amplitude and frequency towards merger; as it does so it sweeps across the sensitive bandwidth of advanced GW interferometric detectors.

For such well-modeled signals, the most efficient signal detection method to extract signals from noisy detector data is \emph{matched filtering}, in which a \emph{template}, representing the predicted waveform as a function of time is correlated with the output of a detector \citep{cwh68}. A matched signal will produce an output much greater than that expected for pure noise with an optimal SNR given as:
\begin{equation}
\rho = 2\left [
  \int_{0}^\infty \, \mathrm{d}f\frac{ \left |\tilde h(f)\right|^2 }{S_h(f)} \right]^{1/2}.
\label{eq:snr}
\end{equation}
For well-modelled sources, matched filtering enhances the value of the signal by a factor $\sqrt{n}$, where $n$ is the number of cycles used in the integration. As inspiralling systems approach merger, even though the rest frame GW amplitude will increase, the number of cycles in each frequency bin, $n = f^{2}(df/dt)^{-1}$, gets smaller; therefore the detected signal will decrease. This means that for inspiraling systems, rather than solely base the predicted amplitude of the radiation as a true indicator of the detectability, we include a measure of the observed cycles. The value of $n$ increases with the compactness of the system as it approaches merger and if observed from a frequency of 10 Hz until merger, could produce $n \sim 10^{4}$ cycles -- effectively improving the detectability by a factor of 100. However, to achieve such gains, a GW data-stream would have to be filtered by a large number of templates (of order $\sim 10^{4} - 10^{5})$ in near real time -- the significant challenges in both theoretical modeling and computational efficiency to achieve this can not be underestimated.

One important aspect of well-modeled inspiralling systems is that a detection can be made 10s of seconds before the merger if enough cycles can be detected to boost the SNR \citep{Manzotti_2012,cannon12}. Figure 2 illustrates this concept showing a chirp signal 40s before the merger phase. This scenario could allow a low-latency alert to be sent out to EM facilities as near real-time as possible to catch a prompt EM signature; the combination of EM and GW data in this regime would provide valuable insight into the inner workings of such cataclysmic events.

It is also worth noting that GWs can provide an independent measure of luminosity distance, $d_L$ \citep{shultz_chirp_86}.  During the inspiral phase, the GW strain, and the rate of change of GW frequency are given as
\begin{equation}
\begin{array}{c}
  \hspace{3.5mm}h       \propto \mathcal{M}_z^{5/3} \hspace{0.5mm}f^{2/3}\hspace{0.5mm}d_{L}  \\
  \dot{f} \propto \mathcal{M}_z^{5/3}\hspace{0.5mm}f^{11/3}\hspace{1mm},
\end{array}
\end{equation}
where $\mathcal{M}_z=\left(1+z\right)\mathcal{M}$ is the redshifted chirp mass, $\mathcal{M} = \left( m_{1} m_{2} \right)^{3/5}/( m_{1} + m_{2})^{1/5}$, and $m_1$, $m_2$ are the component masses of the binary.  Therefore, if one can determine the redshift through, for example, a galaxy association, one can measure the redshift-luminosity distance relation independent of the cosmic distance ladder.  A recent
series
of papers has reinvigorated this topic by introducing novel methods for breaking the redshift-chirp mass degeneracy with future GW observations \citep{Messenger2012PhRvL,Taylor2012PhRvDa, Taylor2012PhRvDb,Nissanke2013arXiv, Messenger2014PhRvX}.

Although matched filtering is the optimal strategy for Gaussian, stationary noise, high amplitude transients due to instrumental and environmental artifacts can render GW data to be non-stationary and non-Gaussian. Therefore, one must employ robust methods that can reject instrumental artifacts and retain the true GW events.

One such method is the $\chi^2$ veto that is a powerful consistency test used to reject false alarms \citep{Allen2005PhRvD}. This method uses the fact that the quantity $\rho$ is an integral over all frequencies and therefore not sensitive to the contributions from different frequency regions of the broadband signal. One can split the signal spectrum into $n$ bins of equal SNR contribution, and draw a comparison with the expected value in each bin (based on the model template). A true GW event will have power accumulated approximately equally in each of $n$ bins; a noise glitch will have power unevenly distributed and will yield a large $\chi^2$ value.

\subsection{The detection range and rates of coalescing compact objects}
\label{section_CBC_Ranges}
\begin{table*}
  \caption{The expected observing scenarios for the aLIGO/AdV era based on \citet{LSC_Prospects_aLIGO_2013}. The available detectors are labeled: \textbf{H} aLIGO-Handford; \textbf{L} aLIGO-Livinstone; \textbf{V} AdV. The aLIGO/AdV detectors will be at design sensitivity by 2019. The expected average ranges for NS/NS and NS/BH inspirals are given in Mpc as well as the horizon distances in parenthesis; these are calculated using equation \ref{eq_inspiral_range} along with the sensitivity noise curves for each of the different observing epochs given in \protect\url{https://dcc.ligo.org/LIGO-T1200307} and assuming masses of 1.4$\msoleq$and 10$\msoleq$ for NSs and BHs respectively.  The detection rates are estimated using the calculated horizon distances along with equation (19) of \citet{Kopparapu_2008ApJ} which is valid for horizon distances greater than 50 Mpc; we obtain estimates in agreement with upper range of the plausible estimates given in \citet{Abadie2010CQGra}.
  }
\begin{center}
\begin{tabular}{|c|c|c|c|c|c|c|}
  \hline
  Observing Run & Duration&Network  &  NS/NS Range   & NS/NS Detection  & NS/BH Range   & NS/BH Detection \\
                &  (months)        &&  Mpc (Horizon) &  Rate (yr$^{-1}$) &  Mpc (Horizon) &  Rate (yr$^{-1}$) \\
                 \hline
Sept 2015 (early) & 3        &LH& 81 (183)    &   $<$ 1          &  168 (380)                     &  $<$  1                 \\
2016-17 (mid)  & 6       &LHV& 121 (273)   &     5         &  253 (572)                     &     2                \\
2017-18 (late) & 9       &LHV& 171 (387)   &     20         &  359 (812)                      &   6                  \\
2019-(final)  &  -         &LHV& 197 (445)   &    40          &  410 (926)                      &    12                 \\
  \hline
\end{tabular}
  \label{table_ligo_runs}
  \end{center}
\end{table*}

In the GW domain, detector sensitivity is generally based on the detection range of BNSs --- the most likely events for detection. The
inspiral horizon distance, $D_{\mathrm{H}}$, is the distance to which an optimally orientated and located equal mass binary can be detected with a SNR equal to 8. For a system with reduced mass, $\mathcal{\mu} = \left( m_{1} m_{2} \right)/( m_{1} + m_{2})$, this distance is approximated as \citep{Singer2014ApJ}:

\begin{equation}\label{eq_inspiral_range}
  D_{\mathrm{H}} = \frac{G^{5/6}M^{1/3}\mu^{1/2}}{c^{3/2}\pi^{2/3}\rho}\sqrt{\frac{5}{6} \int^{ f_{U} }_{ f_{L} }
  \frac{ f^{-7/3} }{S_{n}(f) } } \rm{d}f \,,
\end{equation}

\noindent where $G$ is Newton's gravitational constant, $c$ is the speed of light, $M=m_{1}+m_{2}$ is the total of the system masses, $S_{n}(f)$ the power spectral density of the detectors noise curve and $f$ the signal frequency. The lower limiting frequency of the integral, $f_{L}$ is equal to 10Hz for aLIGO; the upper limiting frequency can be approximated by the last stable orbit of a Schwarzchild black hole, $4400[\msoleq / (m_{1} + m_{2})]\,\rm{Hz} $.

\begin{figure}
        \includegraphics[scale = 0.55,bbllx = 0cm,bblly = 1.0cm, bburx = 13.2cm, bbury = 11cm,origin=lr]{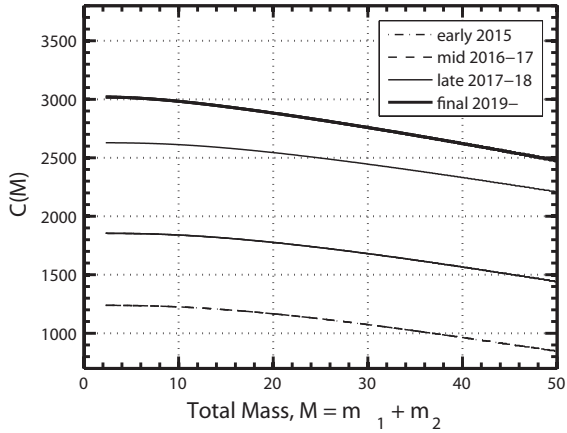}\\
  \caption{To easily approximate the maximum detection ranges for different types of coalescing compact objects, values of C(M) given in equation \ref{eq_inspiral_range_param} are provided by the curves for different values of the combined masses. The curves represent the values of the integral in equation \ref{eq_inspiral_range} for the different values of upper frequency and for the different aLIGO observing scenarios as shown in Table \ref{table_ligo_runs}.}
  \label{fig_CM}
\end{figure}

To calculate approximate values of $ D_{\mathrm{H}} $ a simpler approximation is given by:
\begin{equation}\label{eq_inspiral_range_param}
  D_{\mathrm{H}} = C(M)   \left (  \frac{M}{\msoleq}  \right )^{1/3}
  \left(  \frac{\mu}{\msoleq}   \right)^{1/2}
   \left(  \frac{1}{\rho}   \right )
\end{equation}
\noindent where $C(M)$ gives the value of the integral over $S_{n}(f)$ in equation \ref{eq_inspiral_range} for different $M$; these are calculated for different observing epochs using the sensitivity curves
expected for early aLIGO configurations\footnote{\url{https://dcc.ligo.org/LIGO-T1200307}}. The values of $C(M)$ can then be conveniently read off Figure \ref{fig_CM} for the different observing runs of aLIGO/AdV. Tabulated values of $C(M)$ are provided in Table \ref{table_CM}.

An average range can be obtained by scaling $D_{\mathrm{H}}$ by a factor 2.26 \citep{Singer2014ApJ}. This range assumes a uniform distribution of source sky locations and orientations. A standard figure of merit used by aLIGO/AdV is the \emph{SenseMon Range} which is the average detectable range for two 1.4 $M_{\odot}$ neutron stars \citep{inspiralrange2010}. An additional scaling is given through the association of a GRB with a face on merger which provides an \emph{on-source} time in which to search for a GW event; this increases the sensitivity by a factor of 1.5 \citep[and the corresponding rate of events by a factor 3;][]{Schutz2011CQGra} in comparison with an all-sky/all-time search \citep{Kochanek1993ApJ}; therefore the average orientation average distance of 197 Mpc (see Table 1) becomes 300 Mpc. Thus,
spatially and temporally coincident EM observations enable GW searches to dig deeper into the noise and therefore extend the detection horizon \citep{SuttonPhysRevD2012}.

Table \ref{table_ligo_runs} shows that by 2016-2017  aLIGO/AdV will be accessible to NS/NS inspirals beyond the Coma cluster (100 Mpc). Beyond 2017, with rates of order 20\,yr$^-1$, detections can be expected. The estimates provided in Table \ref{table_ligo_runs} assume a realistic event rate estimates for CBC sources \citep{Abadie2010CQGra}; corresponding numbers that assume plausible pessimistic rate estimates can be obtained by scaling the detection rate estimates down by an order of magnitude. Adopting the latter estimates, there is still a reasonable chance of a NS/NS inspiral and merger detection during 2017.

\section{Gravitational waves from burst sources}
\subsection{Detecting un-modelled burst sources}
\label{section_detection_unmodeled}

Transients that are not well modeled due to their highly complex emissions are also targets for GW detectors; many such \emph{un-modelled bursts} could be associated with LGRBs.

All sky burst searches aim to cast the widest possible net and utilise signal processing algorithms that are as robust as possible; no assumptions are made on the time of arrival, the signals origin or direction. Detection algorithms typically look for signals above a background noise level that are consistent in across multiple detectors; such algorithms often use time-frequency domain methods that look for excesses in time-frequency maps. For example, \verb"X-PIPELINE" combines data from arbitrary detectors in a network and searches for clusters of pixels with energies significantly greater than background \citep{Sutton_2010}. Searches are best employed in networks of detectors using coherent analyses, as described in \S\ref{GW detector networks}. By combining amplitude and phase information from separate detectors in a network, the combined GW signal will increase coherently while the uncorrelated noise can be eliminated. The coherent WaveBurst (\verb"cWB")
is the primary analysis pipeline for identifying burst signals in low-latency \citep{KlimenkoPhysRevD2005}.

\subsection{The detection range and rates of burst events}
\begin{figure}
        \includegraphics[scale = 0.55,bbllx = 0cm,bblly = 1.0cm, bburx = 13.2cm, bbury = 11cm,origin=lr]{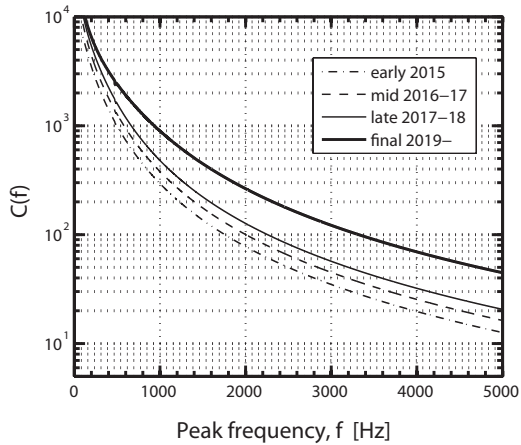}\\
  \caption{To easily approximate the maximum detection ranges for different types of GW burst events, the curves of the function $\mathrm{C}_{\mathrm{B}}$(f) given in equation \ref{eq_burst_range2} are provided for different values of the peak GW frequency. The curves represent the first two components of equation \ref{eq_burst_range} and are shown for different aLIGO observing scenarios as shown in Table \ref{table_ligo_runs}.}
  \label{fig_CB}
\end{figure}

For un-modelled burst sources, the detection strategies are independent of waveform morphology. Therefore, an effective sensitive range\footnote{This range is analogous to the Sensemon range for binary neutron stars.} for a narrowband source can be estimated by considering the total energy emitted in GWs assuming a peak emission frequency, $f_{0}$, for a given SNR $\rho$ \citep{sutton}:
\begin{equation}\label{eq_burst_range}
  D_{\mathrm{Eff}} \approx \left (   \frac{G}{2 \pi^{2} c^{3}  } \right)^{1/2}    \left (   \frac{ 1 } {  S(f_{0}) f_{0}^{2}  }     \right)^{1/2} \left (   \frac{  E_{\rm{GW}}  } { \rho^{2}  }     \right)^{1/2}
\end{equation}
\noindent  One can determine a convenient approximation of $D_{\mathrm{Eff}}$:
\begin{equation}\label{eq_burst_range2}
   D_{\mathrm{Eff}}  \approx C_{\mathrm{B}}(f_{0})  \left (   \frac{  E_{\rm{GW}}  } { \rho^{2}  }     \right)^{1/2},
\end{equation}

\noindent for which, as in \S\ref{section_CBC_Ranges}, values of $C_{B}$ can be derived using the projected sensitivity noise curves for different epochs of observation for aLIGO. Values of $C_{B}$ can be read from Figure \ref{fig_CB} for a given $f_{0}$; tabulated values of $C_{B}$ are provided in Table \ref{table_CB}.

Although there is significant uncertainty in these estimations, as will be discussed later in \S \ref{GRB searches}, such approximations can provide constraints on the global parameters of burst populations such as GRBs.

\section{GW Searches from EM Triggers }
\label{section_gw_from_em}
Gravitational wave interferometers typically continuously collect data from all sky directions. Opportunities exist for both close-to-real-time follow-ups of EM events as well as archival searches. In comparison with all-sky searches using just GW data from an entire science run (of the order of months), an EM triggered search can be conducted over a much smaller time window and sky location. There are a couple of significant advantages with this approach:

\begin{itemize}
  \item A good sky-location enables a search on a portion of the sky with a known antenna pattern sensitivity; this information can allow one to improve the estimation of the GW source parameters.

   \item  The \emph{on-source} data is a window of data taken a short interval before and after the EM trigger time\footnote{Typically large enough to take into account time delays between a GW signal and the onset of the EM emission}. The statistical significance of a GW event in this data segment is determined through a comparison with \emph{off-source} data taken in a period surrounding the on-source window (to represent the noise properties of the on-source segment). The EM trigger time places tighter temporal constraints on the on-source window in comparison with an all-sky all-time search pipeline (\S7.1); a lower number of higher amplitude noise artifacts (non-stationary background noise) will be expected in a smaller interval. As shown by equation \ref{eq_far} this also allows the SNR threshold to be lowered.
\end{itemize}

Numerous archival searches have been carried out using first generation instruments using events such as GRBs \citep{Abadie2012ApJ, Abadie2012ApJa,AasiPhysRevLett2014} and activity from galactic magneters \citep{Abadie2011ApJ}. Although these searches have all produced null results in terms of GW detections, they have enabled the detection procedures for the advanced detector era to be refined as well as providing scientific results. The scientific outcomes of these studies provide an insight into the type of multi-messenger science that could be achieved through the greater detection ranges available in the advanced era. We will describe a few of these below.

\subsection{GRB searches}
\label{GRB searches}
A number of searches have been conducted using LIGO data for coincident GRB events \citep{Abadie2012ApJ, Abadie2012ApJa,Aasi2013PhRvD,2008PhRvD..77f2004A,Predoi2012JPhCS}. The recent GRB search of \citet{Abadie2012ApJ} used 154 GRBs observed during the LIGO and Virgo science runs of 2009-2010 and used both modeled and un-modeled searches in a time-window around the recorded time of the GRB and from the same directions on the sky. For unmodeled bursts, the \verb"X-PIPELINE" method was used to conduct a coherent search, assuming an optimistic emission in GWs of order $10^{-2} M_{\odot}$ and peak emission frequencies of 150 Hz and 300 Hz.
Modeled searches were conducted on the sample of short duration GRBs by combining the data coherently and using template banks corresponding to probable parameters for coalescing systems of NSs and/or BHs \citep{Harry2011PhRvD}, also  yielding exclusion distances --- the distance
beyond which the source must be to avoid detection. The median exclusion distances were 17~Mpc at 150~Hz for the unmodeled search and 16~Mpc for the modeled.

While no GW events were found, none of the observed GRBs fell within the exclusion distance; the closest to date was the \emph{ll}GRB 980425 at 36 Mpc ($z \sim 0.0085$). However, in the advanced detector era, null detections will yield exclusion distances useful to constrain models. For example, two \emph{ll}GRBs observed by \emph{Swift} were at 145 Mpc (GRB 060218) and 264 Mpc (GRB 100316D). Such distances mean that some of the more extreme emission models can be put to the test using GW data.

As discussed earlier in \S 4, following the collapse of a massive star, long-lived ($\sim 10$--1000~s) GW bursts may be produced from rotational instabilities in the protoneutron star remnant \citep{Corsi:2009,Piro_Thrane_2012,Piro_Ott_2011} or in the resulting accretion disk \citep{Putten2008ApJ,Piro_Pfahl_2007}. In either case, the signal is expected to be narrowband with a slowly evolving frequency.

Specialised searches for long-lived GW transients associated with GRBs were conducted but have yielded no candidate detections \citep{Aasi2013PhRvD}. Sensitivity studies suggest that advanced detectors could detect such signals at distances of ~44 Mpc \citep{Thrane_PhysRevD_2014}. There are significant theoretical uncertainties, but the rate of long-lived GW bursts may be sufficiently high for detections with advanced detectors \citep{Piro_Thrane_2012}. Electromagnetic counterparts might include jet-powered type II supernovae, a luminous red nova-like event, or an "un-nova" \citep{Piro_Thrane_2012}.

\subsection{Individual GRB searches}
Gravitational wave searches based on the short-hard GRBs 051103 and 070201 were able to provide some insight into the hosts and emission mechanisms. In the case of GRB 051103, GW data supported evidence that this event was a giant flare of a Soft Gamma-ray Repeater (SGR) \citep{Ofek2006ApJ, Frederiks2007AstL,Hurley_2010}. Triangulation by the inter-planetary network (IPN\footnote{The IPN are a group of gamma-ray burst satellites used to localize GRBs and SGRs through comparison of the arrival times of the events: see
\url{https://heasarc.gsfc.nasa.gov/docs/heasarc/missions/ipn.html}}) suggested that the bright short hard GRB 051103 was in the nearby M81 galaxy (3.6 Mpc). Whether it was from an SGRB (its duration was 0.17~s) or an SGR giant flare was uncertain \citep{Ofek2006ApJ,Hurley_2010}. The energy release, $\sim 5 \times 10^{48}$ ergs assuming it occurred in M81, is a factor of ten times brighter than the brightest SGR giant flare observed \citep[SGR 1806-20;][]{Hurley_2005,Hurley_2010}. Given a typical SGRB energy release of $\sim 10^{50}$ ergs, for a SGRB origin to be compatible, the event could have been a background event to M81 or one would need to invoke a fainter population of short-hard GRBs \citep{Lipunov2005GCN, Hurley_2010}.

Follow up GW searches were performed for both modeled (assuming an inspiraling coalescing binary compact object) and unmodeled bursts (assuming events such as an associated star-quake in a magnetar) \citep{Abadie2012ApJa}. Only the former signal would have been detectible at the distance of M81 \citep{Levin2011MNRAS,Zink2012PhRvD}; the analysis and null result exclude a binary neutron star merger in M81 as the progenitor with 98\% confidence. If the event occurred in M81, the analysis supports the hypothesis of an SGR giant flare producing GRB 051103, which is therefore the most distant extragalactic magnetar observed. Similarly, the study of GRB 070201 \citep{Abbott2008ApJ} observed in M31, provided evidence that this burst did not result from a BNS merger from M31 and is likely to be an SGR giant flare. Given our understanding of SGR giant flares from our own Galaxy, it is statistically unlikely that both GRBs 051103 and 070201 were extragalactic SGR giant flares \citep{Chapman_2009}; hence it is likely that one or both are classical SGRBs from background galaxies.

\section{EM follow-up of GW Triggers}
There is no doubt that low-latency optical and radio follow-ups of GRB triggers revolutionised the field through the discovery of afterglows in the optical and radio \citep{Costa1997Natur,bloom1999Natur}. Similarly, the combination of GW emissions, with complementary EM observations would revolutionise the domain of transient phenomena.

One of the main challenges in achieving this will be the source localisations of the order 100s of deg$^{2}$ \citep{wen10,Fairhurst2011CQGra,chuqi12,LSC_Prospects_aLIGO_2013,Singer2014ApJ,Essick2015ApJ}. Although the nearly omnidirectional GW sensitivity would enable the detection of close EM sources that may be otherwise missed because of their beamed emissions, the large error regions make coordinated followups particularly challenging.

\subsection{The GW detection pipeline}
The main objective of the GW detection pipeline is to identify the most statistically significant GW triggers in the data stream, determine the most probable sky positions and relay the information to partner EM observational facilities as fast as possible -- the general strategy was previously referred to as LOOC-UP \footnote{LOOC-UP stands for \emph{Locating and Observing Optical Counterparts to Unmodeled Pulses} after a pilot study in 2009. We note that this strategy also now encompasses modeled or well predicted sources.} \citep{Kanner_LOOKUP_2008CQGra,Shawhan2012SPIE}. The advanced detector era will see significant improvements in speed; and when a forth detector comes on line, coordinate reconstruction. The basic processes involved in sending out GW triggers to EM partners can be generalised as follows:
\begin{description}
\item[Low-latency data analysis] For well modeled CBC signals, matched filtering (see \S 7.1) is applied to the data using a bank of templates; these are based on the most probable ranges of source parameters e.g. component masses, inclination angles etc. Events above a defined SNR are recorded as triggers. Unmodeled burst searches are also conducted using techniques that are designed to detect a wide range of signals.
  \item[Position reconstruction] Timing triangulation using the differences in the arrival times at each detector in a network can localise the source on the sky \citep{Fairhurst2009NJPh}. At the expense of speed, tighter confidence regions can be determined through more time intensive methods such as coherent analysis. The latter would be beneficial for the optical or radio follow-ups of GRB afterglows.
 \item[Host Galaxy Identification]
      As the positional errors are typically larger than the FoV of most EM instruments (typically tens of square degrees), the probability of a successful follow-up can be improved by using catalogues of nearby galaxies and globular clusters to apply statistical weight on individual tiles (typically 0.4 \degree $\times$ 0.4 \degree ) of an error box \citep{NuttallSutton2010PhRvD,Fan2014ApJ,Bartos2014}. We note that the final aLIGO detection horizon will extend to regions beyond which typical galaxy catalogues have good completeness. Additionally, sources with large galactic offsets could prove problematic \citep{Tunnicliffe2014MNRAS}.
  \item[False Alarm Rate Estimation]  The statistical significance of a GW trigger is given through its FAR already discussed in \S \ref{section_far}. The FAR will identify high significance events that should be considered for follow-up. The FAR represents the average rate at which detector noise fluctuations create false positives with an equal or greater value than the detection statistic or SNR. The rate of background triggers it typically estimated by applying a number of artificial time-shifts of varying durations to the data streams of different detectors in a network around the time of the event -- the time shifts remove any correlations from possible GW signals. By sampling different alignments of the statistical fluctuations a measure of the background rate is obtained that sets the value of the FAR around any GW trigger. A typical FAR threshold adopted to send out alerts during O1 is around 1 event each month of livetime\footnote{The time at which all GW detectors in a network are collecting data}.
  \item[Send out VOEvent] To rapidly communicate the information required by EM facilities for follow-up the VOEvent\footnote{\url{http://www.ivoa.net/documents/VOEvent/}} standard will be adopted \citep{Williams2012SPIE}. This is recognised as the standard syntax for fast dissemination of machine-readable information on astrophysical transients. There are currently different implementations of the VOEvent Transport Protocol\footnote{\url{http://www.ivoa.net/documents/Notes/VOEventTransport/}} that
  have been
  adopted by NASA and ESA space based observatories including {\it Swift} and {\it Fermi } and will be used by the Square Kilometer Array pathfinder telescopes, The Low Frequency Array (LOFAR), ASKAP and MeerKAT. The technical content of a VOEvent alert sent out by aLIGO/AdV for a CBC event should include estimates of the FAR (in Hz), chirp mass the maximum distance (in Mpc); for burst events. content will include central frequency, duration and an estimate of the energy fluence at Earth. Rather than a singular RA/Dec position, the sky position of a GW source will be provided by way of a probability sky map which can be multimodal and non-Gaussian.
\end{description}
\subsection{Communicating GW triggers for EM follow-up}
\label{sec:commtriggers}
If the search pipelines find a candidate signal it is recorded in the GW Candidate event Database (\verb"GraceDB"\footnote{\url{https://gracedb.ligo.org/}}). If its FAR is above threshold, a series of VOEvents are issued. The initial VOEvent will contain only basic information such as the event time, FAR and the GW detectors that have recorded the event. Subsequent VOEvents will contain the information discussed above including skymaps which will provide the probability that the event came from a particular region of sky. The VOEvent will contain a link to the sky map provided in the \verb"HEALPix"\footnote{The acronym stands for \textbf{H}ierarchical \textbf{E}qual \textbf{A}rea iso\textbf{L}atitude \textbf{Pix}elation of a sphere. In this format, all pixels cover equivalent surface areas over a spherical surface \url{http://healpix.sourceforge.net}} format. The first skymap will be a rapid localisation skymap determined by the \verb"BAYESTAR"\footnote{BAYESian TriAngulation and Rapid localisation} pipeline \citep{Singer2014ApJ}. This localisation information can be available within 10s of seconds after detection \citep{Singer_BAYSTAR_2015}.  After further analysis (of order hours) refined full parameter estimation skymaps will be provided using the more rigorous but computationally demanding stochastic samplers in the \verb"LALINFERENCE" pipeline\footnote{\url{https://www.lsc-group.phys.uwm.edu/daswg/projects/lalsuite.html}} that utilises detailed estimates of masses and spins \citep{Berry2015ApJ}.

\begin{figure}
        \includegraphics[scale = 0.9,origin=ll]{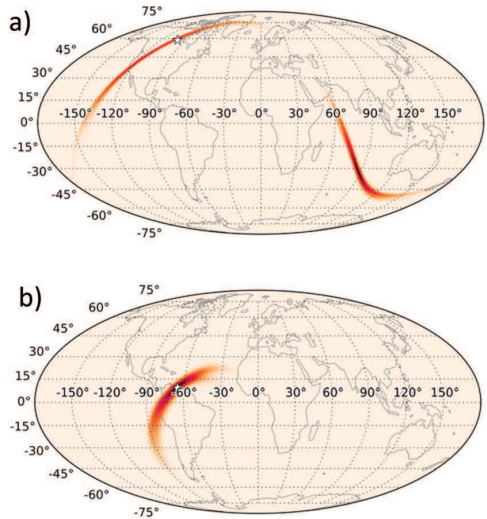}\\
  \caption{Typical GW source skymaps expected from science runs between 2015-2017. The maps are Mollweide projections in geographical coordinates and show: a) two degenerate arcs totalling 820 deg$^{2}$ (event $\sharp$10405) and b) a single elongated arc of 692 deg$^{2}$ (event $\sharp$790258). Both events have a network SNR of 12.7 and the true location of the events are shown by stars. The skymaps are taken from the website repository \protect\url{http://www.ligo.org/scientists/first2years/}.}
  \label{fig_skymaps}
\end{figure}

The morphology of the skymaps are dependent on the location of the source in the sky relative to the GW detector networks antenna pattern function. Some of the probability maps will consist of a single elongated arc which can cover several hundred square degrees, whilst others consist of two or more degenerate arcs. The degeneracy is a result of the two detector networks limited sensitivity to source polarisation \citep{Schutz2011CQGra,Klimenko2011PhRvD}. Figure \ref{fig_skymaps} shows two example skymaps\footnote{Skymaps are taken from the website repository \protect\url{http://www.ligo.org/scientists/first2years/}} typical of that expected from the period 2015-17 which will consist of a 2 detector network and later a third AdV at lower sensitivity (around 36 Mpc range as compared to around 100 Mpc for the aLIGO instruments). The plot shows both a single mode and a bimodal skymaps which will occur in almost equal numbers during this run.

We note that for 2-detector detections during this period both the \verb"BAYESTAR" and stochastic sampler pipelines are expected to produce compatible localisation regions \citep{Singer2014ApJ, Berry2015ApJ}. For the case of 2016-17 with 3-detectors in operation, triggers in all 3 instruments can provide confidence regions of 10s of degrees, although this will occur in less than 17\% of events \citep{Singer2014ApJ}. If AdV records an SNR less than 4, as \verb"BAYESTAR" only considers triggers above SNR=4, it will ignore the third instrument; in this case the stochastic sampler could provide an improved estimate, with up to 50\% smaller area, although within hours latency rather than seconds. By 2019, with aLIGO and AdV running at design sensitivity, up to 25\% of coalescing binary sources are expected to be localised within 20 deg$^{2}$ \citep{LSC_Prospects_aLIGO_2013}.

To fully exploit the scientific promise of rapid GW triggered follow-ups the signal processing will have to be conducted as close to real time as possible (low-latency). This a tremendously complex task and is highly computationally demanding. A number of pipelines have been proposed and tested \citep{first_low_latency_inspiral, mbta,  cannon12}; this has been a particular focus for Australian facilities \citep{ jing,spiir}. Pipelines are presently able to make detections in under a minute \citep{Comm_AlexUrban_2015}, but the effort to get this down to as low as possible will continue throughout the GW multi-messenger era.

\section{Australia's role in gravitational wave astronomy}
\begin{table*}
  \caption{The properties of a selection of the Australian instruments with MoUs in place for aLIGO/AdV follow-ups
  [1] \citet{Tingay2013PASA}  ; [2] \citet{Murphy2013PASAa}; [3] \citet{Tinney04iris2:a}  ;[4] \citet{Keller2007PASA};[5] \citet{Coward2010PASA};  [6] \protect\url{http://goto-observatory.org/} ; [7] \citet{Lennarz2013}; [8] \citet{Acharya_2013APh,Bartos_2014MNRAS}.
  $\flat$ Approximated using Fig. 5 of \citet{Funk2013}. $\dagger$ Sensitivity in survey mode based on \cite{Bartos_2014MNRAS}. Exposure time includes an estimate of the required slewing times to tile a 1000 deg$^{2}$ area using convergent pointing mode.}
\begin{center}
  \begin{tabular}{lllllll}
\hline
\hline
Instrument         &Field-of-view        &Energy       &  Sensitivity    & Exposure         & Response to           & Ref\\
                   &           & range       &         &  Time                    & GW trigger           &   \\
\hline
\hline
MWA                &610\,deg$^{2}$@150\,MHz & 80–-300 MHz  &  10\,mJy                 & 30m                     &   $<10$ secs      & [1] \\
ASKAP (VAST)       &30\,deg$^{2}$@1.4\,GHz  & 700\,MHz-1.8\,GHz& 50\,$\mu$Jy                  &    10s seconds    & mins                        & [2] \\
\hline

AAT          &7arc\,min$^{2}$         &   NIR (J band)  &     22 mag               &  1hr             &  ToO                     & [3]\\
SkyMapper          &5.7 deg$^{2}$ &  Visible (R band) & 21 mag                    &  100s         &  1 - few mins                     & [4]\\
Zadko              &0.15 deg$^{2}$  &  Visible (R band)& 21 mag                  &  180s           &  40s - mins                    & [5]\\
GOTO (Phase 1)      &18 deg$^{2}$  & Visible (R band)& 21 mag                   &   5m             &  mins                      & [6]\\
GOTO (Phase 2)       &36-72 deg$^{2}$  & Visible (R band) & 21 mag                    &  5m             &  mins                       & [6]\\
\hline
H.E.S.S                &15\,deg$^{2}$         & 0.05 -- 20 TeV   &  $6 \times 10^{-8}$ @25GeV   & 1000s  &  $>$30secs        & [7] \\
CTA                 &6 -- 8\,deg$^{2}$    & 0.03 -- 100 TeV &  $6 \times 10^{-9}$@ 25GeV $\flat$  & 1000s  &   20--60secs         & [8] \\
CTA (Survey mode)$\dagger$    &$\sim$1000\,deg$^{2}$  & 0.03 -- 100 TeV &  $6 \times 10^{-8}$@ 25GeV  & 1000s  &   20--60secs         & [8]\\
\hline
\end{tabular}
  \label{table_mous}
  \end{center}
\end{table*}

\subsection{The first follow-up program: 2009-2010}
\label{section_LOOCUP}
The first EM follow-up of GW triggers was performed during 2009--2010 \footnote{This was implemented during two observing periods: Dec. 17, 2010 -- Jan. 8, 2010; Sept. 2 -- Oct. 20, 2010} using the low-latency pipelines cWB, Omega and MBTA \citep[see][]{LSC_Prompt-EM_FuP_2012A&A}. GW data from the LIGO/Virgo network was calibrated and sent to the LIGO computing centre at Caltech for analysis within a minute. Although triggers were generated within 6 minutes, additional manual checks were performed to further varify the data quality and conditions at each detector site -- these latter steps extended the total latency to around 10-30 minutes for each alert. As mentioned in \S \ref{section_LOOCUP}, the strategy in this pilot study was referred to as LOOC-UP \citep{Kanner_LOOKUP_2008CQGra,Shawhan2012SPIE}.

A total of ten EM instruments were employed for LOOK-UP including \emph{Swift}, LOFAR, ROTSE, TAROT, QUEST, the Liverpool Telescope, PTF, and Pi of the Sky; Australian participation was provided in the optical through \emph{SkyMapper} \citep{Keller2007PASA} and the \emph{Zadko Telescope} \citep{Coward2010PASA}. Both instruments responded to GW triggers at a rate of around 1 per week, with 9 and 5 tiles per trigger respectively; in total 8 alerts were followed up \citep{LSC_SearchOpFuP_2014ApJS}. The main latency bottleneck during LOOK-UP was the manual checks on data quality and conditions. To allow alerts to be sent out significantly faster, automation was highlighted as an important prerequisite for coincident detection in the advanced detector era.

\subsection{Multi-messenger astrophysics during the advanced detector era}

The number of Australian facilities with involvement has increased for the advanced detector era. In addition to Zadko and SkyMapper, a number of other instruments have MoUs with the aLIGO/AdV event follow-up program. The full complement is given in Table \ref{table_mous}, along with their relevant specifications.

In the following sections, starting from low-energy observational instruments up to high energy, we discuss these different facilities and their potential contribution towards the multi-messenger era. As discussed in \S
\ref{sec:commtriggers}, the  greatest challenge that will face EM facilities will be contending with the large error regions which could often consist of two or more degenerate arcs - we can not be certain of the exact error regions we will have to overcome. We can however consider two epochs in the following:

\begin{description}
  \item[Early Epoch]
  This epoch includes the early and mid observing runs from 2015--2017 as given in Table 1. The median error regions will be
  in the range
  230--500 deg$^{2}$ \citep{Singer2014ApJ} - we conservatively adopt the larger value of 500 deg$^{2}$ for our approximations. Near the end of this epoch, as AdV joins the 2 aLIGO detectors, one could expect to observe less than 12\% of sources within 20 deg$^{2}$ \citep{LSC_Prospects_aLIGO_2013}, but it is safe to assume that the vast majority of the expected small sample of detections will have error regions of order 100s of deg$^{2}$.

  \item[Late Epoch]During this epoch, aLIGO and AdV will be approaching design sensitivity. One can now expect of order 10--30\% of the detections to be localised within 20\,deg$^{2}$ \citep{LSC_Prospects_aLIGO_2013}. \citet{Chu_inprep} have shown that assuming a three detector aLIGO/AdV network 100\% of sources can be localised within 50 deg$^{2}$ -- we will therefore conservatively adopt this value.  We note that the inclusion of KAGRA in 2018-19 could improve the situation in terms of localisation; \citet{Chu_inprep} further show that including this detector to expand the aLIGO/AdV network will allow 100\% of sources to be localised to within 30 deg$^{2}$.
\end{description}

\subsection{The radio domain}
\label{section_radio}
\subsubsection{Radio Facilities for follow-ups of GW events}
Australian investment in radio facilities and infrastructure has been complemented in recent years by advances in high-speed computing. These new instruments promise a rich era of transient detection by virtue of their wide field-of-view (FoV), high sensitivity and the ability to respond from sub-seconds up to within a minute. Two Australian facilities have signed MoUs with the aLIGO/AdV Event Follow-up program: the MWA \citep[][]{Tingay2013PASA}, and ASKAP \citep[][]{Johnston2007}.

\begin{description}
  \item[The Murchison Widefield Array]\hspace{4cm} The MWA is a low-frequency radio telescope operating between 80 and 300 MHz and located at the Murchison Radio-astronomy Observatory in Western Australia \citep{Tingay2013PASA}. The very large FoV of 610 deg$^{2}$ at 150 MHz and the use of electronic steering make this facility well suited for GW followups.  The MWA can start collecting data within 10\,s of receiving a GW announcement, and additional strategies can be used to survey larger FoVs  at reduced sensitivity if needed \citep{Chu_inprep}. Processing at the start of aLIGO/AdV operation should produce results within 24 hours; this latency could eventually be reduced to less than 1 hour.
\item[Australian Square Kilometre Array Pathfinder] ASKAP consists of an array of 36 $\times$ 12\,m diameter antennas with phased-array feeds in Western Australia. The array can cover an instantaneous FoV of 30 deg$^{2}$, with a resolution of 10-30$^{''}$, 300 MHz bandwidth, and a frequency range of 0.7 to 1.8 GHz. Early science is expected to start in mid-2016. The ASKAP survey for Variables and Slow Transients, VAST \citep{Murphy2013PASAa}, is a survey science program that will
conduct both custom surveys and run commensally with other survey observations. The VAST pipeline will operate on an imaging cadence of 5–-10 seconds at the fastest, down to cadences of minutes depending on the available super-computing resources. Repeated observations of selected fields can allow longer cadences up to hours -- months. Once completed, ASKAP will operate in autonomously mode with ToO response times of order minutes.
\end{description}

Another project that also will have the capability to perform EM follow-ups in the future will be the ASKAP survey for transients on timescales shorter than the correlator integration time.  The Commensal Realtime ASKAP Fast Transient (CRAFT) Survey \citep{jp} performs exactly this task with a commensal survey for fast ($<$5\,s) transient sources, with ASKAP. The extragalactic burst detected by \citet{Lorimer_2007}, with a 30\,Jy pulse of 5\,msec width, provided the first hint of the existence of a previously unknown class of astronomical objects waiting to be discovered. The CRAFT objective is to use the large FoV made available by ASKAP (30 deg$^2$), combined with its excellent sensitivity and resolution, to provide a uniquely capable instrument for both the detection of fast transients and for providing accurate locations to a few arc-seconds of those events. 

\subsubsection{Coordinated radio observations of GW triggers}
For follow-ups in the radio band, the wide FoV of both MWA and ASKAP will be well suited to cover the large GW error region rapidly. The implementation of VOEvent triggering and the fast response times of both these instruments will have great benefits for prompt low-latency follow-ups. Once initial localisation has been achieved other radio telescopes such as the Australia Telescope Compact Array (ATCA) will be valuable for further follow-up. The ATCA has a broadband backend \citep[CABB;][]{Wilson2011} and a rapid response capability through the its Target of Opportunity and NAPA programs.

During the early epoch, the larger FoV of MWA will be well suited for low-latency follow-ups as the large GW error region to be surveyed quickly \citep{Chu_inprep}; in fact, the delays of the low-latency GW analysis may end up dominating the timeline for the MWA, and could limit the types of signals that can be seen. If there is sufficient significance, then a prompt GW alert from before a NS merger could allow MWA to get on-source and prove any association between these events and FRBs. Other than FRBs, MWA will be sensitive to any prompt, coherent emission processes that could accompany SGRBs. A particular advantage of MWA's low-frequency bandpass is that it  any signal will be further delayed through dispersion as it propagates through the ionised intergalactic/interstellar medium. The advantages of this strategy for low-latency follow-ups are clear  and have been discussed in \citet{Chu_inprep}; this additionally adds astrophysical information about the host galaxy and the intergalactic medium \citep[e.g.,][]{Ioka2003, Macquart2013}.

At shorter wavelengths, transient sources that could be accessible by ASKAP could include synchrotron radiation produced through ejected material being accelerated by a magnetar wind (\S 3.2) or through the reverse shock (\S 3.3).  For ASKAP, early follow-ups will only be possible during the late epoch with error regions in the 10s of deg$^{2}$. Although prompt follow-up observations of early engine activity of GW triggers will be challenging for ASKAP during the early epoch, the 30 deg$^2$ FoV can provide good coverage of GW error regions for later time follow-ups. In the GHz regime surveyed by ASKAP there have been observations of late time radio afterglow components from GRBs \citep[see for example][]{Fong2014ApJ} of order hours after the burst. The FoV of ASKAP means that this instrument could return to the same field multiple times to capture the early onset of the light curve to constrain properties of the merger and the local environment.

The observing strategy for CRAFT is to detect any dispersed transient in the total power signal (which is sensitive to the whole 30 deg$^2$ FoV of the telescope) and, after detection, download the raw data from a circular buffer for correlation offline with high temporal resolution. Such a system is compatible with searches based on external triggers from GW detections, if the telescope happened to be pointing in the correct direction. The CRAFT project is 100\% commensal and would be running continuously during all observations. For such a scheme to be successful the GW trigger would need to be communicated to the ASKAP telescope before the circular buffer was over-written; the current specification of the buffer is for 8GB DIMMs, which provides a 40\,second buffer. However the FoV of the ASKAP telescopes mean that such a detection is possible but not likely, and CRAFT is mostly likely to contribute with high time resolution observations during follow-up. The dispersion delay for a signal with a  DM of a few 100 at 700MHz (corresponding with a aLIGO range of 200-450 Mpc), compared to the arrival of the GW, would be of order 2-3 seconds. Even by the late epoch we could not expect such triggering speeds on ASKAP, but for the lower frequency MWA or SKA-low, a longer dispersion delay ($\sim$ 40s @150MHz) would prove valuable for low-latency follow-ups.

The inclusion of the multi-messenger capability to detect and locate very short time scale signals will be an important and unique contribution from the Australian astronomical community. Signals on sub-second timescales would be expected from coherent emission processes at the frequencies covered by ASKAP \citep{clm}, therefore would represent the direct detection of the GW event, not that of the following `fireball'. Recent analysis argues that these will be detectable out to very high redshifts \citep{Lorimer2013MNRAS,Macquart2013}; however the origins and actual physics involved are still so unsure all these arguments are purely conjectural and the answers will come from observations.

Event candidates detected by the ASKAP and MWA pipelines will eventually be distributed in near-real time using the VOEvent standard. The faster an EM counterpart can be communicated to the wider community the greater the opportunity for observations at higher energies when optical/x-ray counterparts may still be detectable.

The significance of any apparent counterpart will have to undergo evaluation for false coincidences. The transient surveys carried out my MWA and ASKAP will be invaluable in this regard. For example, transients observed by the MWA can be evaluated by using a background rate of transient/variable objects \citep{Bell2014MNRAS}. Such rates can be determined though observations of the sky spanning thousands of deg$^{2}$ over many cadences (minutes to years). In addition, for MWA sets of high-quality reference images taken as part of the GaLactic and Extragalactic All-sky MWA (GLEAM) survey \citep{Wayth2015PASA} can act as an important sky template for the study of transient radio sources.  Overall, the rate of astrophysical radio transients is rather low compared to the optical sky \citep[e.g.,][]{Metzger2015,Rowlinson2015}, so that although care must be taken to eliminate instrumental artifacts \citep{Frail2012}, false coincidences will be rare and follow-up effort can be allocated accordingly.

ASKAP and MWA will be detecting and archiving large amounts of transient data with core use of such data for multiwavelength/multi-messenger follow-up searches for counterparts. Therefore Australian radio facilities can also make a contribution to supplying data for archival GW follow-up searches. Such transient searches will follow the same procedures as that of the burst searches for GRB triggers outlined in \S \ref{GRB searches}. GW data streams will be routinely archived allowing early searches around the time of EM triggers, followed by broader archival searches. An archival search can allow one to dig deeper into the GW data stream as an EM trigger provides information of both the sky location and the time of the event. As shown in \S \ref{section_gw_from_em} the FAR will increase with a longer on-source time window, making timing information important. One potential problem is that the timing differentials for different emission mechanisms will have to be well understood; at present, for most sources the expected emissions in the EM domain are quite uncertain. A particular challenge will be to set up automated classification algorithms to catalogue different category of source \citep[e.g.][]{Richards2011,Farrell2015}.

\subsection{The Optical domain}
\label{section_optical}
\subsubsection{Coordinated observations of GW triggers with Optical telescopes}
In the optical, both deep, wide-field instruments and rapidly slewing robotic instruments will have an important role to play. Australia has 4 facilities that are registered as EM partners to aLIGO/AdV: SkyMapper and Zadko conducted follow-ups during the initial LIGO program (2009-2010). The Australian facilaties will be expanded to include the Anglo-Australian Telescope and a new telescope dedicated to GW follow-up, GOTO. We provide a snapshot of these facilities below:

\begin{figure*}
        \includegraphics[scale = 0.7]{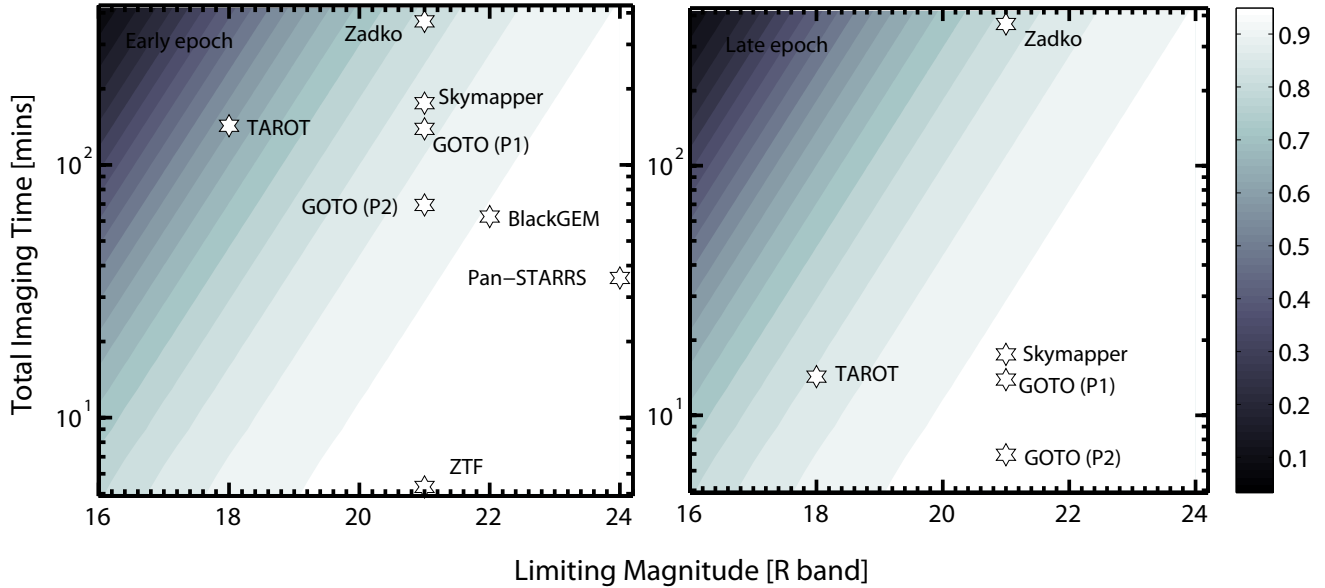}\\
  \caption{A density plot of coincident GW-Optical detection efficiency to recover a SGRB (fading) optical afterglow in the imaging time versus telescope limiting magnitude plane. This plot, adapted from \citet{Coward2014}, shows the Australian optical instruments that have MoUs in place for aLIGO/AdV follow-ups. The total imaging time is the product of the number of tiles required to cover a uniform GW error box for a particular instruments FoV and exposure time. The efficiency, shown by the shaded regions is calculated by considering an optical afterglow luminosity function for SGRBs coupled with limiting magnitude and total imaging time of each instrument. We show results for two scenarios: early epoch (lhs: 500 deg$^2$) and late epoch (rhs: 50 deg$^2$). The Australian facilities Zadko and SkyMapper as well as GOTO (Phase 1, P1 and Phase 2, P2 which will include a second instrument in Australia), Pan-STARRS, BlackGEM and ZTF; three facilities expected to perform with high efficiency in follow-ups during the advanced detector era -- their imaging [time/limiting magnitude] combinations result in their performance being far better the assumed parameter space shown for the late epoch. The efficiencies can be scaled by the expected detection rates and other caveats related to follow-up. We note that GOTO (both P1 and P2) and SkyMapper can make an important contributions to the follow-up program in both epochs. Zadko can make a niche contribution during the latter stages of the advanced detector era as the error regions and detection rates improve.}
  \label{fig_mm_followup}
\end{figure*}

\begin{description}
\item[The Anglo-Australian Telescope (AAT) \hspace{5mm}] AAT is a 4m telescope located at Siding Spring Observatory in NSW, Australia. The AAT has a broad instrument suite, spanning low to high resolution single-object and multi-object optical spectroscopy, as well as near-infrared (NIR) imaging and spectroscopy. The use of optical fibres allows its Two Degree Field (2dF) multi-object system to obtain up to 392 spectra simultaneously from objects within a 2 deg$^2$ FoV. In terms of co-ordinated observations on GW targets, the smaller FoV of the AAT means that the most profitable scenario would be through followup observations of already localised EM signatures. Spectroscopy could be performed if the AAOmega and HERMES instruments were available; both these instruments are fed by the 2dF. This latter scenario would require a delay of up to 1 hour to allow for counterpart confirmation and instrument fibre reconfiguration \citep{Comm_AAO_2015}. Short exposures could be performed without guiding; longer exposures would require guiding but could be achieved using just 2 fibres (one guide and one object fibre). NIR imaging can also be conducted using the IRIS2 instrument \citep[][]{Tinney04iris2:a}; this allows for imaging over a 7 arcmin$^2$ FoV, long-slit spectroscopy and multi-object spectroscopy.

\item[The GW Optical Transient Observer (GOTO)] GOTO is a
proposed network of robotic wide-field
($\sim36$--72~deg$^2$)
optical telescopes to be situated at La Palma, in the Canary Is., and a yet-to-be-determined Australian site. Phase 1 of the project (denoted here as P1), supported primarily by Monash and Warwick Universities (as well as Leicester, Sheffield and Armagh universities in the UK) will deploy a prototype with 18~deg$^2$ FOV (half that of the full-scale instrument) beginning in late 2015, to demonstrate the feasibility of
the approach.
The full-scale instrument will be capable of surveying the entire sky every night and is intended to trigger on GW alerts in real time. A particular goal is to identify candidate transients rapidly in order to trigger other facilities for deeper photometric follow-up and spectroscopic characterisation. The initial  configuration  will consist of a $\sim$18 deg$^{2}$ FoV array in La Palma, Spain, capable of reaching 21 mag in 5 mins (depending on moon phase). To cover the GW error areas in sufficient time, this initial configuration could image at a shallower 20--21~mag, allowing a few hundred degrees to to be surveyed in around 30\,mins. The initial design is scalable and the final configuration will include a second instrument in Australia (denoted here as phase 2, or P2) with
36--72~deg$^{2}$ instantaneous FoV (the larger value for two domes on each site) allowing rapid coverage of GW error ellipses \citep{Comm_GOTO_2015}.
  \item[The SkyMapper telescope \hspace{25mm}] SkyMapper, located at the Siding Spring Observatory in Australia, is a 1.35 m fully autonomous optical telescope with a 5.7 deg$^2$ FoV and equipped with a 268-million pixel CCD array. Its main role is to carry out the Southern Sky Survey \citep{Schmidt2005AAS,Keller2007PASA}; however, a significant component of the SkyMapper science program involves observations of optical transient phenomena. In particular, the SkyMapper Supernova Search, a low-redshift rolling optical survey commencing in 2015, is expected to discover a wide range of optical transients, including Type Ia supernovae for next generation cosmology. The GW follow-up program will benefit from the team's expertise in transient searches. GW triggers received by SkyMapper will take priority over other observations and images will be processed through the transient detection pipeline already developed for the supernova search. Whenever available, images taken as part of the Southern Sky Survey (2015-2018) will serve as pre-detection template. The significance of any optical counter part will be accessed using coincident rate calculated from the supernova search.

\item[The Zadko Telescope \hspace{35mm}] Zadko is a 1m fully robotic instrument with a 23 arcmin FoV located in Gingin, Western Australia. Along with the TAROT\footnote{\url{http://tarot.obs-hp.fr/tarot}} network of fast response telescopes, this instrument has operated successfully as part of a network (CADOR) undertaking automated optical follow up of {\it Swift}\/ alerts (to m$\approx$21) since 2009. It has a core science theme of photometry of rapid time varying sources and it is the most successful Australian-operated facility for GRB afterglow light-curve studies. For GW follow-up, Zadko will be part of a larger network: the TAROT - Zadko - National Aures Observatory Network (TZA). The TAROT network comprises two identical 25 cm, 1.86 deg$^{2}$ robotic telescopes located at Mt Calern in Southern France and ESO La Silla Observatory, Chile. All TAROT telescopes will share a common operating and data processing system. The Algerian National Observatory (Aures) may be operational from 2017 and will comprise of several 50--60~cm telescopes.
\end{description}

\subsubsection{Coordinated optical observations of GW triggers}
The relative sparsity of automated telescopes in the Southern hemisphere implies that instruments such as AAT, GOTO, SkyMapper and Zadko can play an important role in GW follow-ups. This bias has been seen in the sky distribution of \emph{Swift} triggered GRB optical afterglows \citep[see for example, Fig 5. of ][]{Coward2010PASA}. For the case of GRBs, this can hamper the sampling of light curves that last order $\sim$ hours. Hence, both the longitude and latitude of the Australian optical facilities fill a niche space for follow-up.

In Figure \ref{fig_mm_followup} we examine the performance of the larger FoV Australian instruments shown in Table \ref{table_mous} (we have omitted AAT due to its smaller FoV) in terms of obtaining the optical afterglow of a SGRB associated with a NS/NS merger. The general formalism is given in \citet{Coward2014} and considers a measure of the decay of the afterglow with time and a derived luminosity function. The plot is a good illustration of the capabilities of different instruments for rapid response follow-ups. The plot shows that in terms of GW follow-up of SGRBs associated with NS mergers, the first configuration of GOTO (assuming an exposure time of $\sim$ 7.5 mins for Phase 1 and 2 instruments) will be comparative with that of SkyMapper; both are well equipped  for follow-ups and can achieve efficiencies of the order of 80-90\% that of facilities such as BlackGEM \citep{Ghosh2015}, Pan-STARRS \citep{Hodapp2004} and ZTF \citep{Smith_2014SPIE}. Zadko performs well in comparison to the fast response and wider FoV TAROT telescopes because of its sensitivity (TAROTs limit is 18 mag in the R-band and it has a FoV of 3.5 deg$^{2}$).

We note here the coincident detection efficiencies considered in this section ignores a number of other factors including crowded star fields in the Galactic plane and Galactic dust obscuration. Other factors include expertise in dealing with false positives and the ability to apply optimum tiling strategies -- it does however supply a gauge of how well Australian optical facilities can compete in this area.

Follow-up searches for optical r-process kilonova detections could also play an important role in the multi-messenger era. For a source at 200 Mpc the predictions of \citet{Tanaka2013ApJ} suggest that the flux should reach around 21–-23 mag in the optical and 21–-24 mag in the NIR JHK bands (in AB magnitudes). Although the AAT would seem well suited to NIR follow-ups the small FoV of this instrument may make detections difficult for the large error box of a aLIGO/AdV network (10-100 deg$^{2}$ ). However, this instrument could be useful as part of a hierarchical strategy, providing deep follow-up of targets obtained from a larger FoV telescope. The large FoV of SkyMapper is well suited but would require an event with $m <  21$ mag. The final configuration of GOTO ( $m=21$ mag with a large FoV (18-38 deg$^{2}$) suggests this facility could be efficient for follow-up. In fact, dedicated instruments with a wide FoV such as GOTO should play an important role in the multi-messenger era as the first stage of a coordinated follow-up strategy, refining positions for smaller FoV EM instruments.
\subsubsection{Future instruments for GW follow-ups}
Looking towards 2016 and beyond, there are other projects with Australian involvement that can contribute to the GW follow-up program. A new imaging system optimised for low-surface brightness imaging, called Huntsman\footnote{\url{https://www.facebook.com/HuntsmanEye}}, will be based at Siding Spring Observatory. The system consists of an array of Canon telephoto lens based upon the Dragonfly Telephoto Array design \citep{Abraham2014PASP}. With multiple apertures, the system can be automatically configured for shallow imaging over large FoVs or else deep 2x3 deg$^2$ imaging taken with multiple cadences. The response time for a trigger will be a few minutes. The shallow wide-field mode will have an initial field of view of 24 deg$^2$ and will be available from early 2016. It will be upgraded to a field of at least 60 deg$^2$. With 143 mm aperture lenses, the depths in the $r'$-band are approximately 18 AB mag with 9 minute exposures for the shallow field; for 60s exposures, the depth is 16.8 AB mag.

The ``Deeper Wider Faster" project will target simultaneous, fast cadenced observations with optical and radio facilities (Andreoni et al., in prep). The same region of the sky will be observed in the time-domain with the Dark Energy Camera \citep[DECam;][]{Diehl2012, Flaugher2012}, a wide-field optical imager mounted at the prime focus of the Blanco telescope at CTIO, along with the Parkes \citep{Parkes2013PASA} and Molonglo Observatory Synthesis Telescope \citep[MOST;][]{MOST} in Australia. The program takes advantage of the unique ``deep, wide, and fast" capability of DECam reaching a depth of $\sim$\,23.8\,mag ($g$\,filter) in 20\,s and readout time of 17\,s with 62 CCDs covering a FoV of $\sim$\,3\,deg$^2$ per pointing.

Optical and radio data can be processed and analysed in real time to trigger the \textit{Swift} satellite\footnote{Cycle 11 highest priority triggers have been approved for this program} to guarantee fast follow-up of interesting sources in the UV, X--ray, and gamma-ray. These triggers allow other optical facilities to spectroscopically characterise the discovered transients currently via ToO requests.

``Deeper Wider Faster" aims to unveil the optical counterparts to FRBs, along with the discovery of rare and fast (evolving on timescales of seconds to hours) optical transients. Some of these fast transients could be associated with putative GW emitters, some of which have been discussed in \S 3: they include the shock breakouts of nearby core-collapse supernovae \citep[e.g.,][]{Nakar2010}, kilonovae \citep{MetzgerBerger2012ApJ, Tanvir2013Natur}, GRB prompt/early optical emission \citep{Vestrand2014Sci, Fox2003}, and "orphan" GRBs \citep{Ghirlanda2015}. Some models  \citep[e.g.,][]{Falcke_Rezzolla2014AA} argue that FRBs themselves can generate GW radiation.

The program has the capability to identify and reject contaminants in the search for EM counterparts to GWs, such as distant supernovae, stellar flares, tidal disruption events or uncatalogued Active Galactic Nuclei. The project is being led by Swinburne University and is setting up an MoU with the LIGO/Virgo GW collaboration to undertake EM follow-up .

\subsection{Ground-Based Follow-ups in Gamma-rays}
\label{section_VHE}
\subsubsection{Very high energy gamma-ray telescopes for the advanced GW detector era}
At gamma-ray energies from the ground, follow-up observations are possible through Imaging Atmospheric Cherenkov Telescopes (IACTs). These instruments are able to detect gamma-ray photons in the few 10's of GeV to 100 TeV range. They operate by imaging the very short (nanosecond duration) flashes of Cherenkov radiation that result from cascades of relativistic charged particles (known as {\em air-showers}) produced when very high-energy gamma-rays strike the earths atmosphere. A particular target for these instruments will be gamma-rays from SGRBs which are expected as a result of the $>$GeV emission recorded by {\it Fermi}-LAT (see \S 4.1). There are two such facilities with active Australian participation: H.E.S.S. and CTA. A key feature of these telescopes is their huge instantaneous collection area ($> 10^4$\,m$^2$). Flux sensitivities at least a factor 1000 times better than {\it Fermi}-LAT can therefore be achieved over short observations (seconds to hours) in the $\sim$20 to 100 GeV energy range where GRBs are likely to be detected from the ground \citep{Funk2013}.

\begin{description}
\item[The High Energy Stereoscopic System (H.E.S.S.)]
H.E.S.S.\footnote{\url{http://www.mpi-hd.mpg.de/hfm/HESS/}} is an array of 5 Cherenkov telescopes (with 4$\times$ 12m and one 28m diameter mirrors) located in Namibia for TeV or very-high-energy (VHE) gamma-ray astronomy. H.E.S.S. has been operational since 2004 with the fifth larger telescope joining in 2013. The latter instrument lowered the observable energy range from 100\,GeV to a few 10's of GeV and has a rapid slewing capability improving the mean time to go from a random observation position down to about 30 seconds \citep{Lennarz2013}.
\item[The Cherenkov Telescope Array] \hspace{1.7cm}
\mbox{CTA \citep[][]{Acharya_2013APh} \footnote{\url{https://www.cta-observatory.org}}} is a next generation ground-based instrument that will improve over previous experiments such as H.E.S.S, VERITAS\footnote{\url{http://veritas.sao.arizona.edu/}} and MAGIC\footnote{\url{http://magic.mppmu.mpg.de/}} with increased sensitivity, angular resolution, FoV over a wider energy range. The project will consist of two arrays: a southern hemispheric array focusing on Galactic sources and a northern hemispheric array on extragalactic. These will be formed from Cherenkov telescopes of three different sizes ; large (23m diameter), medium (12m) and small (6m) size telescopes, offering wide area and energy coverage. An Australian collaboration of 6 universities led by the University of Adelaide has committed to this project and will contribute expertise through the analysis of CTA data including contributions to the atmospheric calibration. Access to all levels of CTA data will enable Australian collaboration members to contribute towards the GW follow-up program.
\end{description}
\subsubsection{Coordinated observations of GW triggers at high energy gamma ray}
The capabilities of H.E.S.S. for GW follow-ups has been demonstrated through prompt observations of GRB triggers since 2003 --  one of the prime targets for H.E.S.S., and even more so now with the lower-threshold 28m telescope. The fastest follow-up observation was achieved within 7 mins after the burst \citep[GRB 070621;][]{Aharonian2009A&A}. In addition, one burst GRB060602B was fortuitously in the H.E.S.S. FoV on receipt of the trigger (although this may be a galactic transient) and also GRB 100621A was observed within 10 mins.

The wide FoV of CTA will be highly beneficial for GW follow-up allowing the error region to be tiled reasonably rapidly \citep{Bartos2014MNRAS}. CTA's sensitivity (up to a factor 10 better than H.E.S.S.) is expected to guarantee high statistics studies of GRBs well into the multi-GeV regime on minute-wise timescales \citep{Inoue_2013APh}. The CTA is designed to respond to GW alerts by triggering its lowest-threshold telescopes with an expected response of $\geq$20 to 60\,s \citep{Acharya_2013APh} allowing this instrument to make a contribution towards low-latency follow-ups. For GW sources within 200 Mpc, the highest-energy photons will not be effected by degradation by extragalactic background light; therefore the full array can be triggered.

For CTA data processing for new transient sources can be conducted within 30s of taking the data, thus providing the capacity for rapid alerts for GW search pipelines (as is presently done by GRB satellites). Additionally, online analysis can provide nearly real-time data on detections in the FoV; this would enable interesting sources coincident with a GW event to be scrutinised by lower energy instruments.

The direct detection of air shower particles at ground may also be a fruitful way to detect gamma-rays from
GRBs \citep{Bertou_2005}. Although designed to study the highest energy cosmic-rays, the Pierre-Auger Observatory (PAO\footnote{\url{https://www.auger.org/}}), which has Australian involvement, has considered this technique \citep{Allard_2005}.
Using the detection rates from individual Cherenkov water tanks, a $>$100\,MeV fluence (erg\,cm$^{-2}$) sensitivity just beyond that of the brightest {\it Fermi}-LAT GRBs so far observed may be achieved. To-date there is no MoU agreement with PAO but planned upgrades to PAO may offer new opportunities to pursue this avenue. Finally, the High Altitude Water Cherenkov (HAWC) gamma-ray telescope has recently been completed. Its high density sampling of air shower particles at over 4000m above sea level is expected to guarantee detection of at least a few GRBs per year in the $>$100\,GeV band based on {\it Fermi}-LAT detections \citep{Abeysekara_2015ApJ}.

\section{Follow-up by neutrino detection}
\label{section_neutrino}
The IceCube detector at the South Pole was completed in December of 2010, and
monitors a cubic kilometre of deep ice with over 5000 photomultipliers, which
detect Cherenkov light emissions from relativistic particles. Neutrinos can
travel to the Earth from vast distances and if they interact near, or in, the detector volume,
the resulting leptons -  muons, electrons
and taus can be detected. These particles will lose energy to particle showers, the daughter
particles of which in turn will radiate Cherenkov light. The signature of a muon
is a track -- the muon may have a range of many kilometres, producing detected
light in many modules along its path through the detector. Electrons will lose their
energy rapidly, in a short distance (of order a few metres), and result in an
approximately spherical pattern of light outflow from the interaction point. In both cases, there is sufficient information in the shape and magnitude of the timing distributions at the modules
to allow for a reconstruction of the event arrival direction and energy; muons
are resolvable to better than one degree, and cascades to approximately 10-20 degrees.

In the first few years of full operation, IceCube has opened a new observation window on the Universe, with the detection of
high-energy astrophysical neutrinos \citep{i3_HESE2yr,i3_HESE3yr,i3_jakob,i3_ChrisNuMu}. These appear as a excess of events relative to expectations for atmospheric neutrinos, which are the background events made when cosmic rays interact with the Earth's atmosphere.
The highest energy events observed are around 1-2 PeV, and these are the most certain astrophysical events. For lower energy events, each has a probability of being an astrophysical signal relative to the background expectations, and, over many analyses,
the equivalent of about 100 events are thought to be astrophysical. The most definitive events have energies in the hundreds of TeV range and above, with several events observed beyond 1 PeV. Possible sources for these neutrinos include particle acceleration environments in our own galaxy, and in other galaxies. The deep reach of neutrinos means that IceCube can probe particle acceleration processes out to redshifts of 1 and beyond. The ongoing goal of IceCube is to determine the sources and production mechanisms of the observed neutrinos, and  finding a neutrino signal in coincidence with another messenger would yield
critical information about the neutrino sources.

IceCube operates in full-sky coverage mode at near 100\% uptime, making it ideal for followup studies of other messengers such as GW sources. If a GW signal is discovered, the already-collected IceCube data from the discovery time may be retrieved and
checked to see if any neutrinos were in coincidence. To this end, IceCube has formalised agreements with LIGO/Virgo for the joint analysis of data. The first analyses have been published, covering periods of joint operation from 2007-2010 \citep{ILV}. This joint analysis assigns significance to
GW and neutrino events separately, and then these significances are combined. In this first analysis, no significant correlations are seen. Work is underway to analyse the full detector data that now exists, and to prepare for next-generation GW detectors coming online.

\section{Summary}
The current network of interferometric GW detectors offer the very real prospect of providing an entirely new avenue for understanding the Universe. It is anticipated that a key capability to maximise scientific return from the detector network will be the ability to detect EM counterparts for GW sources.

One of the most promising EM sources for co-ordinated GW observations are GRBs. It is widely assumed that the progenitors of these events are cataclysmic sources, such as the collapse of massive stars and coalescing systems of compact binaries. These events are also detection targets for the GW domain. In this review we focused on GRBs to consider some of the multi-messenger scenarios that may be possible with GWs.

Discovery possibilities are numerous and highly uncertain at this time. Coordinated GW observations of short duration GRBs could yield conclusive proof of a connection with compact binary mergers. A low-latency detection of a coalescing compact object 10s of seconds before the merger could allow fast response telescopes to be on-source at the time of the merger and thus observe the prompt and early emissions (\S \ref{section_cbc}). Such a scenario could be the key to unlocking mysteries such as the mechanisms behind long lived X-ray plataus (\S \ref{section_magnetar}) and the observed VHE gamma-ray emissions (\S \ref{section_prompt}) and to test if any connection with fast radio bursts exists (\S \ref{section_prompt} \& \S \ref{section_magnetar}). If instabilities exist in the collapse of massive stars, the enhanced GW emissions could be detected from a local population of low-luminosity GRBs and coupled with EM or neutrino observations of the burst and an associated supernova (\S \ref{section_LGRBs}). Many other coordinated EM observations are possible with GW triggers at both early and late times. We should also be prepared for serendipitous discoveries.

While searches for such counterparts present technical challenges, past achievements in detecting counterparts for other types of transients in large error regions are encouraging \citep{SingerPTF2015ApJ}. Teams of observers with wide-field instruments across the EM spectrum are already preparing for EM-followups. Different follow-up techniques are being tested, including sophisticated tiling strategies and machine-learning approaches for screening of candidate counterparts.

At the same time new wide-field radio facilities in Australia including ASKAP, MWA, and eventually SKA will offer an expanded ability to detect transient sources in very large fields (\S \ref{section_radio}). These developing capabilities, coupled with Australia's geographic advantage in terms of access to a large fraction of the Southern sky implies that ground-based followup in both the optical and radio seems particularly promising. Certainly, the geographic location is proven for telescopes like AAT, SkyMapper and Zadko and in the future, GOTO (\S \ref{section_optical}) can also capitalise. The energetics of GW sources suggest that Australian involvement in both high-energy gamma  (\S \ref{section_VHE}) and neutrino observations  (\S \ref{section_neutrino}) could offer unique capabilities. Although extremely challenging, participation in this new era has the potential to place Australia at the forefront of arguably the most exciting discoveries for 21$^{\mathrm{st}}$ century astronomy.
%
\section{ACKNOWLEDGMENTS}
A number of people have aided this paper through valuable discussions and by providing information on the different featured EM facilities: we particularly wish to thank, Chris Lidman (AAO); Paul O’Brien and Jim Hinton (CTA); Danny Steeghs (GOTO); Shami Chatterjee (VAST); Jeff Cooke (``Deeper Wider Faster"). EJH acknowledges support from a UWA Research Fellowship. DC is supported by an Australian Research Council Future Fellowship (FT100100345).  PDL is supported by the Australian Research Council Discovery Project (DP140102578). Part of this research was conducted by the Australian Research Council
Centre of Excellence for All-sky Astrophysics (CAASTRO), through project number CE110001020. DLK was supported by NSF grant AST-1412421. The authors thank Marica Branchesi, the assigned reviewer for the LIGO Scientific Collaboration, for conducting a thorough review of the manuscript which included a number of insightful suggestions. We also acknowledge the anonymous referee who highlighted a number of areas that have benefited from improved clarity. This is LIGO document number LIGO-P1500153.

\begin{appendix}
\begin{table*}
  \caption{The values of $C(M)$ as given in equation \ref{eq_inspiral_range} for the sensitivities corresponding with the different observation runs of aLIGO/AdV. These data can be interpolated and used to calculate estimates of the gravitational wave detection ranges of coalescing compact objects. }
\begin{center}
\begin{tabular}{|c|c|c|c|c|}
  \hline
 Total Mass & $C(M)$ &  $C(M)$  & $C(M)$  & $C(M)$    \\
 $M_{\odot}$ & Early (2015) & Mid (2016-17)  & Late (2017-18)  & Final (2019-)    \\
 \hline
2.80  &   1238.40   &   1854.09  &   2628.37  &   3018.87  \\
5.30  &   1236.61   &   1852.10  &   2626.13  &   3012.00  \\
7.80  &   1232.09   &   1847.10  &   2620.54  &   2998.58  \\
10.30  &   1224.33   &   1838.61  &   2611.16  &   2980.39  \\
12.80  &   1213.39   &   1826.70  &   2598.12  &   2958.47  \\
15.30  &   1199.54   &   1811.76  &   2581.94  &   2933.95  \\
17.80  &   1182.75   &   1793.83  &   2562.77  &   2906.95  \\
20.30  &   1164.21   &   1774.21  &   2542.02  &   2879.18  \\
22.80  &   1143.22   &   1752.18  &   2518.99  &   2849.44  \\
25.30  &   1120.75   &   1728.78  &   2494.77  &   2818.98  \\
27.80  &   1096.06   &   1703.21  &   2468.57  &   2786.63  \\
30.30  &   1070.66   &   1677.00  &   2441.96  &   2754.20  \\
32.80  &   1044.19   &   1649.75  &   2414.49  &   2721.04  \\
35.30  &   1018.20   &   1622.97  &   2387.69  &   2688.89  \\
37.80  &   987.34   &   1591.08  &   2355.95  &   2650.97  \\
40.30  &   959.26   &   1561.87  &   2327.02  &   2616.47  \\
42.80  &   931.83   &   1533.09  &   2298.60  &   2582.61  \\
45.30  &   900.83   &   1500.17  &   2266.14  &   2543.94  \\
47.80  &   871.91   &   1469.01  &   2235.42  &   2507.29  \\
50.30  &   839.90   &   1433.88  &   2200.72  &   2465.84  \\
  \hline
\end{tabular}
  \label{table_CM}
  \end{center}
\end{table*}

\end{appendix}

\begin{appendix}

\begin{table*}
  \caption{The values of $C_{\rm{B}}(f)$ as given in equation \ref{eq_burst_range} for the sensitivities corresponding with the different observation runs of aLIGO/AdV. These data can be interpolated and used to calculate estimates of the gravitational wave detection ranges of burst sources of different peak frequencies. }
\begin{center}
\begin{tabular}{|c|c|c|c|c|}
  \hline
 Peak Frequency & $C_{\rm{B}}(f)$ $\times 10^{3}$ &  $C_{\rm{B}}(M)$ $\times 10^{3}$ & $C_{\rm{B}}(M)$ $\times 10^{3}$ & $C_{\rm{B}}(M)$ $\times 10^{3}$   \\
 Hz &  Early (2015) &  Mid (2016-17)  &  Late (2017-18)  & Final (2019-)    \\
 \hline
100.00  &   6.52   &   8.64  &   10.52  &   12.17  \\
200.00  &   3.48   &   4.40  &   5.39  &   6.50  \\
300.00  &   2.11   &   2.69  &   3.33  &   4.33  \\
400.00  &   1.41   &   1.80  &   2.25  &   3.16  \\
500.00  &   0.99   &   1.28  &   1.60  &   2.42  \\
600.00  &   0.74   &   0.95  &   1.19  &   1.91  \\
700.00  &   0.56   &   0.73  &   0.91  &   1.54  \\
800.00  &   0.44   &   0.57  &   0.72  &   1.27  \\
900.00  &   0.36   &   0.46  &   0.58  &   1.06  \\
1000.00  &   0.29   &   0.38  &   0.48  &   0.90  \\
1100.00  &   0.25   &   0.32  &   0.40  &   0.77  \\
1200.00  &   0.21   &   0.27  &   0.34  &   0.66  \\
1300.00  &   0.18   &   0.23  &   0.29  &   0.58  \\
1400.00  &   0.16   &   0.20  &   0.25  &   0.51  \\
1500.00  &   0.14   &   0.18  &   0.22  &   0.45  \\
1600.00  &   0.12   &   0.15  &   0.20  &   0.40  \\
1700.00  &   0.11   &   0.14  &   0.17  &   0.36  \\
1800.00  &   0.10   &   0.12  &   0.16  &   0.32  \\
1900.00  &   0.09   &   0.11  &   0.14  &   0.29  \\
2000.00  &   0.08   &   0.10  &   0.13  &   0.26  \\
2100.00  &   0.07   &   0.09  &   0.12  &   0.24  \\
2200.00  &   0.06   &   0.08  &   0.11  &   0.22  \\
2300.00  &   0.06   &   0.08  &   0.10  &   0.20  \\
2400.00  &   0.05   &   0.07  &   0.09  &   0.19  \\
2500.00  &   0.05   &   0.06  &   0.08  &   0.17  \\
2600.00  &   0.05   &   0.06  &   0.08  &   0.16  \\
2700.00  &   0.04   &   0.06  &   0.07  &   0.15  \\
2800.00  &   0.04   &   0.05  &   0.07  &   0.14  \\
2900.00  &   0.04   &   0.05  &   0.06  &   0.13  \\
3000.00  &   0.03   &   0.05  &   0.06  &   0.12  \\
\hline
\end{tabular}
  \label{table_CB}
  \end{center}
\end{table*}
\end{appendix}


\begin{thebibliography}{255}
\expandafter\ifx\csname natexlab\endcsname\relax\def\natexlab#1{#1}\fi

\bibitem[{{Aartsen} {et~al.}(2015{\natexlab{a}}){Aartsen}, {Abraham},
  {Ackermann}, {et~al.}}]{i3_ChrisNuMu}
{Aartsen}, M.~G., {Abraham}, K., {Ackermann}, M., {et~al.} 2015{\natexlab{a}},
  Phys. Rev. Lett., 115, 081102

\bibitem[{{Aartsen} {et~al.}(2014{\natexlab{a}}){Aartsen}, {Ackermann},
  {Adams}, {Aguilar}, {Ahlers}, {Ahrens}, {Altmann}, {Anderson}, {Arguelles},
  {Arlen}, \& et~al.}]{i3_HESE3yr}
{Aartsen}, M.~G., {Ackermann}, M., {Adams}, J., {Aguilar}, J.~A., {Ahlers}, M.,
  {Ahrens}, M., {Altmann}, D., {Anderson}, T., {Arguelles}, C., {Arlen}, T.~C.,
  \& et~al. 2014{\natexlab{a}}, Physical Review Letters, 113, 101101

\bibitem[{{Aartsen} {et~al.}(2014{\natexlab{b}}){Aartsen}, {Ackermann},
  {Adams}, {et~al.}}]{ILV}
{Aartsen}, M.~G., {Ackermann}, M., {Adams}, J., {et~al.} 2014{\natexlab{b}},
  \prd, 90, 102002

\bibitem[{{Aartsen} {et~al.}(2015{\natexlab{b}}){Aartsen}, {Ackermann},
  {Adams}, {et~al.}}]{i3_jakob}
{Aartsen}, M.~G., {Ackermann}, M., {Adams}, J., {et~al.} 2015{\natexlab{b}}, \prd, 91, 022001

\bibitem[{{Aasi} {et~al.}(2014){Aasi}, {Abadie}, {Abbott}, {Abbott},
  {et~al.}}]{LSC_SearchOpFuP_2014ApJS}
{Aasi}, J., {Abadie}, J., {Abbott}, B.~P., {Abbott}, R., {et~al.} 2014, \apjs,
  211, 7

\bibitem[{{Aasi} {et~al.}(2013{\natexlab{a}}){Aasi}, {Abadie}, {Abbott},
  {et~al.}}]{Aasi2013PhRvD}
{Aasi}, J., {Abadie}, J., {Abbott}, B.~P., {et~al.} 2013{\natexlab{a}}, \prd,
  88, 122004

\bibitem[{{Aasi} {et~al.}(2015){Aasi}, {Abbott}, {Ligo Scientific
  Collaboration}, {et~al.}}]{aLIGO_2015}
{Aasi}, J., {Abbott}, B., {Ligo Scientific Collaboration}, {et~al.} 2015,
  Classical and Quantum Gravity, 32, 074001

\bibitem[{Aasi {et~al.}(2014)Aasi, Abbott, Abbott,
  {et~al.}}]{AasiPhysRevLett2014}
Aasi, J., Abbott, B.~P., Abbott, R., {et~al.} 2014, Phys. Rev. Lett., 113,
  011102

\bibitem[{{Aasi} {et~al.}(2013{\natexlab{b}})}]{LSC_Prospects_aLIGO_2013}
{Aasi}, J. {et~al.} 2013{\natexlab{b}}, astro-ph: 1304.0670

\bibitem[{{Abadie} {et~al.}(2010{\natexlab{a}}){Abadie}, {Abbott}, {Abbott},
  {et~al.}}]{inspiralrange2010}
{Abadie}, J., {Abbott}, B.~P., {Abbott}, R., {et~al.} 2010{\natexlab{a}},
  astro-ph/1003.2481

\bibitem[{{Abadie} {et~al.}(2011){Abadie}, {Abbott}, {Abbott},
  {et~al.}}]{Abadie2011ApJ}
{Abadie}, J., {Abbott}, B.~P., {Abbott}, R., {et~al.} 2011, \apjl, 734, L35

\bibitem[{{Abadie} {et~al.}(2012{\natexlab{a}}){Abadie}, {Abbott}, {Abbott},
  {et~al.}}]{first_low_latency_inspiral}
{Abadie}, J., {Abbott}, B.~P., {Abbott}, R., {et~al.} 2012{\natexlab{a}}, \aap, 541, A155

\bibitem[{{Abadie} {et~al.}(2012{\natexlab{b}}){Abadie}, {Abbott}, {Abbott},
  {et~al.}}]{Abadie2012ApJ}
{Abadie}, J., {Abbott}, B.~P., {Abbott}, R., {et~al.} 2012{\natexlab{b}}, \apj, 760, 12

\bibitem[{{Abadie} {et~al.}(2012{\natexlab{c}}){Abadie}, {Abbott}, {Abbott},
  {et~al.}}]{Abadie2012ApJa}
{Abadie}, J., {Abbott}, B.~P., {Abbott}, T.~D., {et~al.} 2012{\natexlab{c}},
  \apj, 755, 2

\bibitem[{{Abadie} {et~al.}(2012{\natexlab{d}}){Abadie}, {Ligo Scientific
  Collaboration}, {Virgo Collaboration}, {et~al.}}]{LSC_Prompt-EM_FuP_2012A&A}
{Abadie}, J., {Ligo Scientific Collaboration}, {Virgo Collaboration}, {et~al.}
  2012{\natexlab{d}}, \aap, 539, A124

\bibitem[{{Abadie} {et~al.}(2010{\natexlab{b}})}]{Abadie2010CQGra}
{Abadie}, J. {et~al.} 2010{\natexlab{b}}, Classical and Quantum Gravity, 27,
  173001

\bibitem[{{Abbott} {et~al.}(2008{\natexlab{a}}){Abbott}, {Abbott}, {Adhikari},
  {et~al.}}]{Abbott2008ApJ}
{Abbott}, B., {Abbott}, R., {Adhikari}, R., {et~al.} 2008{\natexlab{a}}, \apj,
  681, 1419

\bibitem[{{Abbott} {et~al.}(2008{\natexlab{b}}){Abbott}, {Abbott}, {Adhikari},
  {et~al.}}]{Abbott2008PhRvL}
{Abbott}, B., {Abbott}, R., {Adhikari}, R., {et~al.} 2008{\natexlab{b}}, Physical Review Letters, 101, 211102

\bibitem[{{Abbott} {et~al.}(2008{\natexlab{c}}){Abbott}, {Abbott}, {Adhikari},
  {et~al.}}]{2008PhRvD..77f2004A}
{Abbott}, B., {Abbott}, R., {Adhikari}, R., {et~al.} 2008{\natexlab{c}}, \prd, 77, 062004

\bibitem[{{Abeysekara} {et~al.}(2015){Abeysekara}, {Alfaro}, {Alvarez}, \&
  the{HAWC Collaboration}}]{Abeysekara_2015ApJ}
{Abeysekara}, A.~U., {Alfaro}, R., {Alvarez}, C. e.~a., \& the{HAWC
  Collaboration}. 2015, \apj, 800, 78

\bibitem[{{Abraham} \& {van Dokkum}(2014)}]{Abraham2014PASP}
{Abraham}, R.~G. \& {van Dokkum}, P.~G. 2014, \pasp, 126, 55

\bibitem[{{Accadia} {et~al.}(2012){Accadia}, {Acernese}, {Alshourbagy},
  {et~al.}}]{virgo2012JInst}
{Accadia}, T., {Acernese}, F., {Alshourbagy}, M., {et~al.} 2012, Journal of
  Instrumentation, 7, 3012

\bibitem[{Acernese {et~al.}(2015)Acernese, Agathos, Agatsuma,
  {et~al.}}]{Acernese_2015}
Acernese, F., Agathos, M., Agatsuma, K., {et~al.} 2015, Classical and Quantum
  Gravity, 32, 024001

\bibitem[{{Acharya} {et~al.}(2013){Acharya}, {Actis}, {Aghajani},
  {et~al.}}]{Acharya_2013APh}
{Acharya}, B.~S., {Actis}, M., {Aghajani}, T., {et~al.} 2013, Astroparticle
  Physics, 43, 3

\bibitem[{{Ackermann} {et~al.}(2010){Ackermann}, {Asano}, {Atwood}, {Axelsson},
  {Baldini}, {et~al.}}]{Ackermann_2010ApJ}
{Ackermann}, M., {Asano}, K., {Atwood}, W.~B., {Axelsson}, M., {Baldini}, L.,
  {et~al.} 2010, \apj, 716, 1178

\bibitem[{{Aharonian} {et~al.}(2009)}]{Aharonian2009A&A}
{Aharonian}, F. {et~al.} 2009, \aap, 495, 505

\bibitem[{{Akerlof} {et~al.}(1999)}]{Akerlof1999Natur}
{Akerlof}, C. {et~al.} 1999, \nat, 398, 400

\bibitem[{{Allard} {et~al.}(2005){Allard}, {Parizot}, {Bertou}, {Beatty}, {Du
  Vernois}, {Nitz}, \& {Rodriguez}}]{Allard_2005}
{Allard}, D., {Parizot}, E., {Bertou}, X., {Beatty}, J., {Du Vernois}, M.,
  {Nitz}, D., \& {Rodriguez}, G. 2005, International Cosmic Ray Conference, 4,
  427

\bibitem[{{Allen}(2005)}]{Allen2005PhRvD}
{Allen}, B. 2005, \prd, 71, 062001

\bibitem[{{Andersson}(1998)}]{Andersson1998ApJ}
{Andersson}, N. 1998, \apj, 502, 708

\bibitem[{{Andersson} \& {Kokkotas}(2001)}]{Andersson2001IJMPD}
{Andersson}, N. \& {Kokkotas}, K.~D. 2001, International Journal of Modern
  Physics D, 10, 381

\bibitem[{Arnaud {et~al.}(2002)}]{Arnaud02}
Arnaud, N. {et~al.} 2002, Phys. Rev. D, 65, 042004

\bibitem[{{Barnard} {et~al.}(2008){Barnard}, {Clark}, \&
  {Kolb}}]{Barnard_IC10X-1_2008}
{Barnard}, R., {Clark}, J.~S., \& {Kolb}, U.~C. 2008, A \& A, 488, 697

\bibitem[{{Bartos} {et~al.}(2014{\natexlab{a}}){Bartos}, {Crotts}, \&
  {Marka}}]{Bartos2014}
{Bartos}, I., {Crotts}, A.~P.~S., \& {Marka}, S. 2014{\natexlab{a}}, ArXiv
  e-prints

\bibitem[{{Bartos} {et~al.}(2014{\natexlab{b}}){Bartos}, {Veres}, {Nieto},
  {Connaughton}, {Humensky}, {Hurley}, {M{\'a}rka}, {M{\'e}sz{\'a}ros},
  {Mukherjee}, {O'Brien}, \& {Osborne}}]{Bartos2014MNRAS}
{Bartos}, I., {Veres}, P., {Nieto}, D., {Connaughton}, V., {Humensky}, B.,
  {Hurley}, K., {M{\'a}rka}, S., {M{\'e}sz{\'a}ros}, P., {Mukherjee}, R.,
  {O'Brien}, P., \& {Osborne}, J.~P. 2014{\natexlab{b}}, \mnras, 443, 738

\bibitem[{{Bartos} {et~al.}(2014{\natexlab{c}}){Bartos}, {Veres}, {Nieto},
  {et~al.}}]{Bartos_2014MNRAS}
{Bartos}, I., {Veres}, P., {Nieto}, D., {et~al.} 2014{\natexlab{c}}, \mnras,
  443, 738

\bibitem[{{Bauswein} {et~al.}(2013){Bauswein}, {Goriely}, \&
  {Janka}}]{Bauswein2013ApJ}
{Bauswein}, A., {Goriely}, S., \& {Janka}, H.-T. 2013, \apj, 773, 78

\bibitem[{{Bell} {et~al.}(2014){Bell}, {Murphy}, {Kaplan},
  {et~al.}}]{Bell2014MNRAS}
{Bell}, M.~E., {Murphy}, T., {Kaplan}, D.~L., {et~al.} 2014, \mnras, 438, 352

\bibitem[{{Berger}(2007)}]{Berger_2007}
{Berger}, E. 2007, \apj, 670, 1254

\bibitem[{{Berger}(2009)}]{Berger_2009ApJ}
{Berger}, E. 2009, \apj, 690, 231

\bibitem[{{Berger} {et~al.}(2013){Berger}, {Fong}, \&
  {Chornock}}]{Berger2013ApJ}
{Berger}, E., {Fong}, W., \& {Chornock}, R. 2013, \apjl, 774, L23

\bibitem[{{Berger} {et~al.}(2005)}]{Berger2005Natur}
{Berger}, E. {et~al.} 2005, \nat, 438, 988

\bibitem[{{Berry} {et~al.}(2015){Berry}, {Mandel}, {Middleton},
  {et~al.}}]{Berry2015ApJ}
{Berry}, C.~P.~L., {Mandel}, I., {Middleton}, H., {et~al.} 2015, \apj, 804, 114

\bibitem[{{Bertou} \& {Allard}(2005)}]{Bertou_2005}
{Bertou}, X. \& {Allard}, D. 2005, Nuclear Instruments and Methods in Physics
  Research A, 553, 299

\bibitem[{{Blair} {et~al.}(2008){Blair}, {Barriga}, {Brooks},
  {et~al.}}]{blair08}
{Blair}, D.~G., {Barriga}, P., {Brooks}, A.~F., {et~al.} 2008, Journal of
  Physics Conference Series, 122, 012001

\bibitem[{{Bloom} {et~al.}(1999){Bloom}, {Kulkarni}, {Djorgovski},
  {et~al.}}]{bloom1999Natur}
{Bloom}, J.~S., {Kulkarni}, S.~R., {Djorgovski}, S.~G., {et~al.} 1999, \nat,
  401, 453

\bibitem[{{Bloom} {et~al.}(2006)}]{Bloom2006ApJ}
{Bloom}, J.~S. {et~al.} 2006, \apj, 638, 354

\bibitem[{{Bouvier} {et~al.}(2011){Bouvier}, {Gilmore}, {Connaughton}, {Otte},
  {Primack}, \& {Williams}}]{Bouvier_2011}
{Bouvier}, A., {Gilmore}, R., {Connaughton}, V., {Otte}, N., {Primack}, J.~R.,
  \& {Williams}, D.~A. 2011, astro-ph/1109.5680

\bibitem[{{Branchesi} {et~al.}(2012){Branchesi}, {Ligo Scientific
  Collaboration}, \& {Virgo Collaboration}}]{Branchesi2012JPhCS}
{Branchesi}, M., {Ligo Scientific Collaboration}, \& {Virgo Collaboration}.
  2012, Journal of Physics Conference Series, 375, 062004

\bibitem[{{Bromberg} {et~al.}(2013){Bromberg}, {Nakar}, {Piran}, \&
  {Sari}}]{Bromberg_2013}
{Bromberg}, O., {Nakar}, E., {Piran}, T., \& {Sari}, R. 2013, \apj, 764, 179

\bibitem[{{Bucciantini} {et~al.}(2009)}]{Bucciantini_2009}
{Bucciantini}, N. {et~al.} 2009, MNRAS, 396, 2038

\bibitem[{{Buskulic} {et~al.}(2010){Buskulic}, {Virgo Collaboration}, \& {LIGO
  Scientific Collaboration}}]{mbta}
{Buskulic}, D., {Virgo Collaboration}, \& {LIGO Scientific Collaboration}.
  2010, Classical and Quantum Gravity, 27, 194013

\bibitem[{{Campana} {et~al.}(2006){Campana}, {Mangano}, {Blustin},
  {et~al.}}]{Campana2006Natur}
{Campana}, S., {Mangano}, V., {Blustin}, A.~J., {et~al.} 2006, \nat, 442, 1008

\bibitem[{{Cannon} {et~al.}(2012){Cannon}, {Cariou}, {Chapman},
  {et~al.}}]{cannon12}
{Cannon}, K., {Cariou}, R., {Chapman}, A., {et~al.} 2012, \apj, 748, 136

\bibitem[{{Cavalier} {et~al.}(2006){Cavalier}, {Barsuglia}, {Bizouard},
  {et~al.}}]{cavalier06}
{Cavalier}, F., {Barsuglia}, M., {Bizouard}, M., {et~al.} 2006, \prd, 74,
  082004

\bibitem[{{Chapman} {et~al.}(2009){Chapman}, {Priddey}, \&
  {Tanvir}}]{Chapman_2009}
{Chapman}, R., {Priddey}, R.~S., \& {Tanvir}, N.~R. 2009, \mnras, 395, 1515

\bibitem[{Chu {et~al.}(2015)Chu, Howell, Rowlinson, {et~al.}}]{Chu_inprep}
Chu, Q., Howell, E., Rowlinson, A., {et~al.} 2015, submitted to MNRAS:
  astro-ph/1509.06876

\bibitem[{Chu {et~al.}(2012)Chu, Wen, \& Blair}]{chuqi12}
Chu, Q., Wen, L., \& Blair, D. 2012, Journal of Physics: Conference Series,
  363, 012023

\bibitem[{{Cook} {et~al.}(1994){Cook}, {Shapiro}, \& {Teukolsky}}]{Cook1994ApJ}
{Cook}, G.~B., {Shapiro}, S.~L., \& {Teukolsky}, S.~A. 1994, \apj, 424, 823

\bibitem[{{Cordes} {et~al.}(2004){Cordes}, {Lazio}, \& {McLaughlin}}]{clm}
{Cordes}, J.~M., {Lazio}, T.~J.~W., \& {McLaughlin}, M.~A. 2004, \nar, 48, 1459

\bibitem[{{Corsi} \& {M{\'e}sz{\'a}ros}(2009{\natexlab{a}})}]{Corsi2009ApJ}
{Corsi}, A. \& {M{\'e}sz{\'a}ros}, P. 2009{\natexlab{a}}, \apj, 702, 1171

\bibitem[{{Corsi} \& {M{\'e}sz{\'a}ros}(2009{\natexlab{b}})}]{Corsi:2009}
{Corsi}, A. \& {M{\'e}sz{\'a}ros}, P. 2009{\natexlab{b}}, ApJ, 702, 1171

\bibitem[{{Costa} {et~al.}(1997){Costa}, {Frontera}, {Heise},
  {et~al.}}]{Costa1997Natur}
{Costa}, E., {Frontera}, F., {Heise}, J., {et~al.} 1997, \nat, 387, 783

\bibitem[{{Coward}(2005)}]{coward_LLGRB_05}
{Coward}, D.~M. 2005, \mnras, 360, L77

\bibitem[{{Coward} {et~al.}(2014){Coward}, {Branchesi}, {Howell}, {Lasky}, \&
  {B{\"o}er}}]{Coward2014}
{Coward}, D.~M., {Branchesi}, M., {Howell}, E.~J., {Lasky}, P.~D., \&
  {B{\"o}er}, M. 2014, \mnras, 445, 3575

\bibitem[{{Coward} {et~al.}(2010){Coward}, {Todd}, {Vaalsta},
  {et~al.}}]{Coward2010PASA}
{Coward}, D.~M., {Todd}, M., {Vaalsta}, T.~P., {et~al.} 2010, \pasa, 27, 331

\bibitem[{{Cutler}(2002)}]{Cutler2002PhRvD}
{Cutler}, C. 2002, \prd, 66, 084025

\bibitem[{{Cutler} \& {Flanagan}(1994)}]{cutler94}
{Cutler}, C. \& {Flanagan}, {\'E}.~E. 1994, \prd, 49, 2658

\bibitem[{{Dai} \& {Lu}(1998)}]{Dai1998}
{Dai}, Z.~G. \& {Lu}, T. 1998, \aap, 333, L87

\bibitem[{{Daigne} \& {Mochkovitch}(2007)}]{Daigne_2007}
{Daigne}, F. \& {Mochkovitch}, R. 2007, \aap, 465, 1

\bibitem[{{Dall'Osso} {et~al.}(2015){Dall'Osso}, {Giacomazzo}, {Perna}, \&
  {Stella}}]{DallOsso2015ApJ}
{Dall'Osso}, S., {Giacomazzo}, B., {Perna}, R., \& {Stella}, L. 2015, \apj,
  798, 25

\bibitem[{{Diehl} \& {Dark Energy Survey Collaboration}(2012)}]{Diehl2012}
{Diehl}, H.~T. \& {Dark Energy Survey Collaboration}. 2012, in American
  Astronomical Society Meeting Abstracts, Vol. 219, American Astronomical
  Society Meeting Abstracts \#219, \#413.05

\bibitem[{{Dimmelmeier} {et~al.}(2008){Dimmelmeier}, {Ott}, {Marek}, \&
  {Janka}}]{Dimmelmeier2008PhRvD}
{Dimmelmeier}, H., {Ott}, C.~D., {Marek}, A., \& {Janka}, H.-T. 2008, \prd, 78,
  064056

\bibitem[{{Duncan} \& {Thompson}(1992)}]{DuncanThomson_92}
{Duncan}, R.~C. \& {Thompson}, C. 1992, ApJL, 392, L9

\bibitem[{{Eichler} {et~al.}(1989{\natexlab{a}}){Eichler}, {Livio}, {Piran}, \&
  {Schramm}}]{Eichler1989Natur}
{Eichler}, D., {Livio}, M., {Piran}, T., \& {Schramm}, D.~N.
  1989{\natexlab{a}}, \nat, 340, 126

\bibitem[{{Elliott} {et~al.}(2014)}]{Elliott2014A&A}
{Elliott}, J. {et~al.} 2014, \aap, 562, A100

\bibitem[{{Essick} {et~al.}(2015){Essick}, {Vitale}, {Katsavounidis},
  {Vedovato}, \& {Klimenko}}]{Essick2015ApJ}
{Essick}, R., {Vitale}, S., {Katsavounidis}, E., {Vedovato}, G., \& {Klimenko},
  S. 2015, \apj, 800, 81

\bibitem[{{Fairhurst}(2009)}]{Fairhurst2009NJPh}
{Fairhurst}, S. 2009, New Journal of Physics, 11, 123006

\bibitem[{{Fairhurst}(2011)}]{Fairhurst2011CQGra}
{Fairhurst}, S. 2011, Classical and Quantum Gravity, 28, 105021

\bibitem[{{Falcke} \& {Rezzolla}(2014)}]{Falcke_Rezzolla2014AA}
{Falcke}, H. \& {Rezzolla}, L. 2014, \aap, 562, A137

\bibitem[{{Fan} {et~al.}(2014){Fan}, {Messenger}, \& {Heng}}]{Fan2014ApJ}
{Fan}, X., {Messenger}, C., \& {Heng}, I.~S. 2014, \apj, 795, 43

\bibitem[{{Fan} {et~al.}(2013){Fan}, {Wu}, \& {Wei}}]{Fan2013PhRvD}
{Fan}, Y.-Z., {Wu}, X.-F., \& {Wei}, D.-M. 2013, \prd, 88, 067304

\bibitem[{{Farrell} {et~al.}(2015){Farrell}, {Murphy}, \& {Lo}}]{Farrell2015}
{Farrell}, S., {Murphy}, T., \& {Lo}, K. 2015, \apj, In press

\bibitem[{Finn(2002)}]{Finn02}
Finn, L.~S. 2002, Phys. Rev. D, 63, 102001

\bibitem[{{Flaugher} {et~al.}(2012){Flaugher}, {Abbott}, {Angstadt},
  {et~al.}}]{Flaugher2012}
{Flaugher}, B.~L., {Abbott}, T.~M.~C., {Angstadt}, R., {et~al.} 2012, in
  Society of Photo-Optical Instrumentation Engineers (SPIE) Conference Series,
  Vol. 8446, Society of Photo-Optical Instrumentation Engineers (SPIE)
  Conference Series, 11

\bibitem[{{Fong} {et~al.}(2014){Fong}, {Berger}, {Metzger},
  {et~al.}}]{Fong2014ApJ}
{Fong}, W., {Berger}, E., {Metzger}, B.~D., {et~al.} 2014, \apj, 780, 118

\bibitem[{{Fox} {et~al.}(2003){Fox}, {Price}, {Soderberg}, {Berger},
  {Kulkarni}, {Sari}, {Frail}, {Harrison}, {Yost}, {Matthews}, {Peterson},
  {Tanaka}, {Christiansen}, \& {Moriarty-Schieven}}]{Fox2003}
{Fox}, D.~W., {Price}, P.~A., {Soderberg}, A.~M., {Berger}, E., {Kulkarni},
  S.~R., {Sari}, R., {Frail}, D.~A., {Harrison}, F.~A., {Yost}, S.~A.,
  {Matthews}, K., {Peterson}, B.~A., {Tanaka}, I., {Christiansen}, J., \&
  {Moriarty-Schieven}, G.~H. 2003, ApJ, 586, L5

\bibitem[{{Frail} {et~al.}(2012){Frail}, {Kulkarni}, {Ofek}, {Bower}, \&
  {Nakar}}]{Frail2012}
{Frail}, D.~A., {Kulkarni}, S.~R., {Ofek}, E.~O., {Bower}, G.~C., \& {Nakar},
  E. 2012, \apj, 747, 70

\bibitem[{{Frederiks} {et~al.}(2007){Frederiks}, {Palshin}, {Aptekar},
  {Golenetskii}, {Cline}, \& {Mazets}}]{Frederiks2007AstL}
{Frederiks}, D.~D., {Palshin}, V.~D., {Aptekar}, R.~L., {Golenetskii}, S.~V.,
  {Cline}, T.~L., \& {Mazets}, E.~P. 2007, Astronomy Letters, 33, 19

\bibitem[{Fruchter {et~al.}(2006)}]{Fruchter_grb_gal}
Fruchter, A.~S. {et~al.} 2006, \nat, 441, 463

\bibitem[{{Fryer} {et~al.}(2002){Fryer}, {Holz}, \& {Hughes}}]{Fryer2002ApJ}
{Fryer}, C.~L., {Holz}, D.~E., \& {Hughes}, S.~A. 2002, \apj, 565, 430

\bibitem[{Fryer \& New(2011)}]{Fryer2011}
Fryer, C.~L. \& New, K.~C. 2011, Living Reviews in Relativity, 14


\bibitem[{{Funk} {et~al.}(2013{\natexlab{b}}){Funk}, {Hinton}, \& {CTA
  Consortium}}]{Funk2013}
{Funk}, S., {Hinton}, J.~A., \& {CTA Consortium}. 2013{\natexlab{b}}, Astroparticle Physics, 43, 348

\bibitem[{{Gao} {et~al.}(2013{\natexlab{a}}){Gao}, {Ding}, {Wu}, {Zhang}, \&
  {Dai}}]{Gao2013}
{Gao}, H., {Ding}, X., {Wu}, X.-F., {Zhang}, B., \& {Dai}, Z.-G.
  2013{\natexlab{a}}, \apj, 771, 86


\bibitem[{{Gehrels} {et~al.}(2008){Gehrels}, {Barthelmy}, {Burrows},
  {et~al.}}]{Gehrels_2008}
{Gehrels}, N., {Barthelmy}, S.~D., {Burrows}, D.~N., {et~al.} 2008, \apj, 689,
  1161

\bibitem[{{Gehrels} {et~al.}(2005){Gehrels}, {Sarazin}, {O'Brien},
  {et~al.}}]{gehrels_2005}
{Gehrels}, N., {Sarazin}, C.~L., {O'Brien}, P.~T., {et~al.} 2005, \nat, 437,
  851

\bibitem[{Gehrels {et~al.}(2004)}]{Gehrels_Swift_2004}
Gehrels, N. {et~al.} 2004, \apj, 611, 1005

\bibitem[{{Gendre} {et~al.}(2013){Gendre}, {Stratta}, {Atteia},
  {et~al.}}]{Gendre2013ApJ}
{Gendre}, B., {Stratta}, G., {Atteia}, J.~L., {et~al.} 2013, \apj, 766, 30

\bibitem[{{Ghirlanda} {et~al.}(2015){Ghirlanda}, {Salvaterra}, {Campana},
  {et~al.}}]{Ghirlanda2015}
{Ghirlanda}, G., {Salvaterra}, R., {Campana}, S., {et~al.} 2015, A\&A, 578, A71

\bibitem[{{Ghosh} \& {Nelemans}(2015)}]{Ghosh2015}
{Ghosh}, S. \& {Nelemans}, G. 2015, Astrophysics and Space Science Proceedings,
  40, 51

\bibitem[{{Giacomazzo} \& {Perna}(2013)}]{Giacomazzo2013ApJ}
{Giacomazzo}, B. \& {Perna}, R. 2013, \apjl, 771, L26

\bibitem[{Greiner {et~al.}(2015)}]{Greiner_2015}
Greiner, J. {et~al.} 2015, Nature, 523, 189

\bibitem[{{Guetta} \& {Della Valle}(2007)}]{GuettaDellaValle_2007}
{Guetta}, D. \& {Della Valle}, M. 2007, \apjl, 657, L73

\bibitem[{{Harry} \& {Fairhurst}(2011)}]{Harry2011PhRvD}
{Harry}, I.~W. \& {Fairhurst}, S. 2011, \prd, 83, 084002

\bibitem[{Hartle(2003)}]{Hartle03}
Hartle, J.~B. 2003, Gravity (San Francisco: Addison and Wesley), pp. 506-7

\bibitem[{{Haskell} {et~al.}(2008){Haskell}, {Samuelsson}, {Glampedakis}, \&
  {Andersson}}]{Haskell2008MNRAS}
{Haskell}, B., {Samuelsson}, L., {Glampedakis}, K., \& {Andersson}, N. 2008,
  \mnras, 385, 531

\bibitem[{Helstrom(1968)}]{cwh68}
Helstrom, C. 1968, International Series of Monographs in Electronics and
  Instrumentation, Vol.~9, Statistical Theory of Signal Detection, 2nd edn.
  (Oxford; New York: Pergamon Press)

\bibitem[{Hild {et~al.}(2011)Hild, Abernathy, Acernese, {et~al.}}]{Hild_2011}
Hild, S., Abernathy, M., Acernese, F., {et~al.} 2011, Classical and Quantum
  Gravity, 28, 094013

\bibitem[{{Hild} {et~al.}(2008){Hild}, {Chelkowski}, \& {Freise}}]{Hild_2008}
{Hild}, S., {Chelkowski}, S., \& {Freise}, A. 2008, astro-ph/0810.0604

\bibitem[{Hild {et~al.}(2010)Hild, Chelkowski, Freise, Franc, Morgado,
  Flaminio, \& DeSalvo}]{Hild_2010}
Hild, S., Chelkowski, S., Freise, A., Franc, J., Morgado, N., Flaminio, R., \&
  DeSalvo, R. 2010, Classical and Quantum Gravity, 27, 015003

\bibitem[{Hjorth(2003)}]{Hjorth_LGRB_SN}
Hjorth, J. 2003, \nat, 423, 847

\bibitem[{{Hodapp} {et~al.}(2004){Hodapp}, {Siegmund}, {Kaiser},
  {et~al.}}]{Hodapp2004}
{Hodapp}, K.~W., {Siegmund}, W.~A., {Kaiser}, N., {et~al.} 2004, in Society of
  Photo-Optical Instrumentation Engineers (SPIE) Conference Series, Vol. 5489,
  Ground-based Telescopes, ed. J.~M. {Oschmann}, Jr., 667--678

\bibitem[{{Hooper} {et~al.}(2012){Hooper}, {Chung}, {Luan}, {Blair}, {Chen}, \&
  {Wen}}]{spiir}
{Hooper}, S., {Chung}, S.~K., {Luan}, J., {Blair}, D., {Chen}, Y., \& {Wen}, L.
  2012, \prd, 86, 024012

\bibitem[{{Howell} \& {Coward}(2013)}]{howell_2013}
{Howell}, E.~J. \& {Coward}, D.~M. 2013, \mnras, 428, 167

\bibitem[{Hulse \& Taylor(1975)}]{Hulse_Taylor_75}
Hulse, R.~A. \& Taylor, J.~H. 1975, Astrophys. J., 195, L51

\bibitem[{{Hurley} {et~al.}(2005){Hurley}, {Boggs}, \& {Smith}}]{Hurley_2005}
{Hurley}, K., {Boggs}, S.~E., \& {Smith}, D.~M.~a. 2005, \nat, 434, 1098

\bibitem[{{Hurley} {et~al.}(2010){Hurley}, {Rowlinson}, {Bellm},
  {et~al.}}]{Hurley_2010}
{Hurley}, K., {Rowlinson}, A., {Bellm}, E., {et~al.} 2010, \mnras, 403, 342

\bibitem[{{IceCube Collaboration}(2013)}]{i3_HESE2yr}
{IceCube Collaboration}. 2013, Science, 342, 1242856

\bibitem[{{IceCube Collaboration} {et~al.}(2006){IceCube Collaboration},
  {Achterberg}, {Ackermann}, {Adams}, {Ahrens}, {Andeen}, {Atlee}, {Baccus},
  {Bahcall}, {Bai}, \& et~al.}]{IC2006APh}
{IceCube Collaboration}, {Achterberg}, A., {Ackermann}, M., {Adams}, J.,
  {Ahrens}, J., {Andeen}, K., {Atlee}, D.~W., {Baccus}, J., {Bahcall}, J.~N.,
  {Bai}, X., \& et~al. 2006, Astroparticle Physics, 26, 155

\bibitem[{{Imerito} {et~al.}(2008)}]{Imerito_08}
{Imerito}, A. {et~al.} 2008, MNRAS, 391, 405

\bibitem[{{Inoue} {et~al.}(2013){Inoue}, {Granot}, {O'Brien}, \& {for the CTA
  Consortium}}]{Inoue_2013APh}
{Inoue}, S., {Granot}, J., {O'Brien}, \& {for the CTA Consortium}. 2013,
  Astroparticle Physics, 43, 252

\bibitem[{{Ioka}(2003)}]{Ioka2003}
{Ioka}, K. 2003, \apjl, 598, L79

\bibitem[{Jaranowski {et~al.}(1998)Jaranowski, Krolak, \&
  Schutz}]{jaranowski_anttenapatfuns_98}
Jaranowski, P., Krolak, A., \& Schutz, B.~F. 1998, Phys. Rev. D, 58, 063001

\bibitem[{{Johnston} {et~al.}(2007){Johnston}, {Bailes}, {Bartel},
  {et~al.}}]{Johnston2007}
{Johnston}, S., {Bailes}, M., {Bartel}, N., {et~al.} 2007, \pasa, 24, 174

\bibitem[{{Kann} {et~al.}(2011){Kann}, {Klose}, {Zhang}, {et~al.}}]{Kann_2011}
{Kann}, D.~A., {Klose}, S., {Zhang}, B., {et~al.} 2011, \apj, 734, 96

\bibitem[{{Kanner} {et~al.}(2008){Kanner}, {Huard}, {M{\'a}rka},
  {et~al.}}]{Kanner_LOOKUP_2008CQGra}
{Kanner}, J., {Huard}, T.~L., {M{\'a}rka}, S., {et~al.} 2008, Classical and
  Quantum Gravity, 25, 184034

\bibitem[{{Keller} {et~al.}(2007){Keller}, {Schmidt}, {Bessell},
  {et~al.}}]{Keller2007PASA}
{Keller}, S.~C., {Schmidt}, B.~P., {Bessell}, M.~S., {et~al.} 2007, \pasa, 24,
  1

\bibitem[{Klimenko {et~al.}(2005)Klimenko, Mohanty, Rakhmanov, \&
  Mitselmakher}]{KlimenkoPhysRevD2005}
Klimenko, S., Mohanty, S., Rakhmanov, M., \& Mitselmakher, G. 2005, Phys. Rev.
  D, 72, 122002

\bibitem[{{Klimenko} {et~al.}(2011){Klimenko}, {Vedovato}, {Drago},
  {et~al.}}]{Klimenko2011PhRvD}
{Klimenko}, S., {Vedovato}, G., {Drago}, M., {et~al.} 2011, \prd, 83, 102001

\bibitem[{{Kobayashi} \& {M{\'e}sz{\'a}ros}(2003)}]{Kobayashi2003ApJ}
{Kobayashi}, S. \& {M{\'e}sz{\'a}ros}, P. 2003, \apj, 589, 861

\bibitem[{{Kochanek} \& {Piran}(1993)}]{Kochanek1993ApJ}
{Kochanek}, C.~S. \& {Piran}, T. 1993, \apjl, 417, L17

\bibitem[{{Kopparapu} {et~al.}(2008){Kopparapu}, {Hanna}, {Kalogera},
  {et~al.}}]{Kopparapu_2008ApJ}
{Kopparapu}, R.~K., {Hanna}, C., {Kalogera}, V., {et~al.} 2008, \apj, 675, 1459

\bibitem[{{Kouveliotou} {et~al.}(1993){Kouveliotou}, {Meegan}, {Fishman},
  {et~al.}}]{Kouveliotou_1993}
{Kouveliotou}, C., {Meegan}, C.~A., {Fishman}, G.~J., {et~al.} 1993, \apjl,
  413, L101

\bibitem[{Kuroda \& the LCGT~Collaboration(2010)}]{KAGRA}
Kuroda, K. \& the LCGT~Collaboration. 2010, Classical and Quantum Gravity, 27,
  084004

\bibitem[{{Lai} \& {Shapiro}(1995)}]{LaiShapiro_1995}
{Lai}, D. \& {Shapiro}, S.~L. 1995, ApJ, 442, 259

\bibitem[{{Lasky}(2015)}]{lasky15pasa}
{Lasky}, P.~D. 2015, \pasa\,in press, arXiv:1508.06643

\bibitem[{{Lasky} \& {Glampedakis}(2015)}]{Lasky2015inprep}
{Lasky}, P.~D. \& {Glampedakis}, K. 2015, in preparation

\bibitem[{{Lasky} {et~al.}(2014){Lasky}, {Haskell}, {Ravi}, {Howell}, \&
  {Coward}}]{Lasky2014PhRvD}
{Lasky}, P.~D., {Haskell}, B., {Ravi}, V., {Howell}, E.~J., \& {Coward}, D.~M.
  2014, \prd, 89, 047302

\bibitem[{{Lennarz} {et~al.}(2013){Lennarz}, {Chadwick}, {Domainko},
  {et~al.}}]{Lennarz2013}
{Lennarz}, D., {Chadwick}, P.~M., {Domainko}, W., {et~al.} 2013,
  astro-ph:1307.6897

\bibitem[{{Levin} \& {van Hoven}(2011)}]{Levin2011MNRAS}
{Levin}, Y. \& {van Hoven}, M. 2011, \mnras, 418, 659

\bibitem[{{Li} \& {Paczy{\'n}ski}(1998)}]{Li1998ApJ}
{Li}, L.-X. \& {Paczy{\'n}ski}, B. 1998, \apjl, 507, L59

\bibitem[{Lidman(2015)}]{Comm_AAO_2015}
Lidman, C. 2015, personal communication

\bibitem[{{Lipunov} {et~al.}(2005){Lipunov}, {Kornilov}, {Kuvshinov},
  {et~al.}}]{Lipunov2005GCN}
{Lipunov}, V., {Kornilov}, V., {Kuvshinov}, D., {et~al.} 2005, GRB Coordinates
  Network, 4206, 1

\bibitem[{{Lorimer} {et~al.}(2007){Lorimer}, {Bailes}, {McLaughlin},
  {Narkevic}, \& {Crawford}}]{Lorimer_2007}
{Lorimer}, D.~R., {Bailes}, M., {McLaughlin}, M.~A., {Narkevic}, D.~J., \&
  {Crawford}, F. 2007, Science, 318, 777

\bibitem[{{Lorimer} {et~al.}(2013){Lorimer}, {Karastergiou}, {McLaughlin}, \&
  {Johnston}}]{Lorimer2013MNRAS}
{Lorimer}, D.~R., {Karastergiou}, A., {McLaughlin}, M.~A., \& {Johnston}, S.
  2013, \mnras, 436, L5

\bibitem[{{Luan} {et~al.}(2011){Luan}, {Hooper}, {Wen}, \& {Chen}}]{jing}
{Luan}, J., {Hooper}, S., {Wen}, L., \& {Chen}, Y. 2011, ArXiv e-prints
  1108.3174

\bibitem[{Lyons {et~al.}(2010)}]{Lyons_2010}
Lyons, N. {et~al.} 2010, MNRAS, 402, 705

\bibitem[{{Macquart}(2007)}]{Macquart_2007}
{Macquart}, J.-P. 2007, \apjl, 658, L1

\bibitem[{{Macquart} {et~al.}(2010){Macquart}, {Bailes}, {Bhat}, , \& {for the
  CRAFT Collaboration}}]{jp}
{Macquart}, J.-P., {Bailes}, M., {Bhat}, N.~D.~R., , \& {for the CRAFT
  Collaboration}. 2010, \pasa, 27, 272

\bibitem[{{Macquart} \& {Koay}(2013)}]{Macquart2013}
{Macquart}, J.-P. \& {Koay}, J.~Y. 2013, \apj, 776, 125

\bibitem[{{Manchester} {et~al.}(2013){Manchester}, {Hobbs}, {Bailes},
  {et~al.}}]{Parkes2013PASA}
{Manchester}, R.~N., {Hobbs}, G., {Bailes}, M., {et~al.} 2013, \pasa, 30, 17

\bibitem[{{Mandel} {et~al.}(2012){Mandel}, {Kelley}, \&
  {Ramirez-Ruiz}}]{Mandel_2012IAUS}
{Mandel}, I., {Kelley}, L.~Z., \& {Ramirez-Ruiz}, E. 2012, in IAU Symposium,
  Vol. 285, IAU Symposium, ed. E.~{Griffin}, R.~{Hanisch}, \& R.~{Seaman},
  358--360

\bibitem[{{Manzotti} \& {Dietz}(2012)}]{Manzotti_2012}
{Manzotti}, A. \& {Dietz}, A. 2012, astro-ph/1202.4031

\bibitem[{Maselli \& Ferrari(2014)}]{Maselli_PhysRevD_2014}
Maselli, A. \& Ferrari, V. 2014, Phys. Rev. D, 89, 064056

\bibitem[{{Messenger} \& {Read}(2012)}]{Messenger2012PhRvL}
{Messenger}, C. \& {Read}, J. 2012, Physical Review Letters, 108, 091101

\bibitem[{{Messenger} {et~al.}(2014){Messenger}, {Takami}, {Gossan},
  {Rezzolla}, \& {Sathyaprakash}}]{Messenger2014PhRvX}
{Messenger}, C., {Takami}, K., {Gossan}, S., {Rezzolla}, L., \&
  {Sathyaprakash}, B.~S. 2014, Physical Review X, 4, 041004

\bibitem[{{Metzger} {et~al.}(2014){Metzger}, {Bauswein}, {Goriely}, \&
  {Kasen}}]{Metzger_2014}
{Metzger}, B.~D., {Bauswein}, A., {Goriely}, S., \& {Kasen}, D. 2014,
  astro-ph/1409.0544

\bibitem[{{Metzger} \& {Berger}(2012)}]{MetzgerBerger2012ApJ}
{Metzger}, B.~D. \& {Berger}, E. 2012, \apj, 746, 48

\bibitem[{{Metzger} {et~al.}(2011){Metzger}, {Giannios}, {Thompson},
  {Bucciantini}, \& {Quataert}}]{Metzger2011MNRAS}
{Metzger}, B.~D., {Giannios}, D., {Thompson}, T.~A., {Bucciantini}, N., \&
  {Quataert}, E. 2011, \mnras, 413, 2031

\bibitem[{{Metzger} {et~al.}(2010){Metzger}, {Mart{\'{\i}}nez-Pinedo},
  {Darbha}, {et~al.}}]{Metzger2010MNRAS}
{Metzger}, B.~D., {Mart{\'{\i}}nez-Pinedo}, G., {Darbha}, S., {et~al.} 2010,
  \mnras, 406, 2650

\bibitem[{{Metzger} {et~al.}(2008){Metzger}, {Piro}, \&
  {Quataert}}]{Metzger_2008}
{Metzger}, B.~D., {Piro}, A.~L., \& {Quataert}, E. 2008, \mnras, 390, 781

\bibitem[{{Metzger} {et~al.}(2015){Metzger}, {Williams}, \&
  {Berger}}]{Metzger2015}
{Metzger}, B.~D., {Williams}, P.~K.~G., \& {Berger}, E. 2015,
  astro-ph/1502.01350

\bibitem[{{Miller}(2005)}]{miller_05}
{Miller}, M.~C. 2005, Astrophys. J. Lett, 626, L41

\bibitem[{Mills.(1981)}]{MOST}
Mills., B.~Y. 1981, Proc. Astron. Soc. Australia, 4, 156

\bibitem[{{Moortgat} \& {Kuijpers}(2005)}]{Moortgat_2005}
{Moortgat}, J. \& {Kuijpers}, J. 2005, in 22nd Texas Symposium on Relativistic
  Astrophysics, ed. P.~{Chen}, E.~{Bloom}, G.~{Madejski}, \& V.~{Patrosian},
  326--331

\bibitem[{Murase {et~al.}(2006)}]{murase_06}
Murase, K. {et~al.} 2006, \apj, 651, L5

\bibitem[{{Murphy} {et~al.}(2013)}]{Murphy2013PASAa}
{Murphy}, T. {et~al.} 2013, \pasa, 30, 6

\bibitem[{{Nakar}(2015)}]{Nakar2015ApJ}
{Nakar}, E. 2015, \apj, 807, 172

\bibitem[{{Nakar} \& {Sari}(2010)}]{Nakar2010}
{Nakar}, E. \& {Sari}, R. 2010, ApJ, 725, 904

\bibitem[{{Narayan} {et~al.}(1992){Narayan}, {Paczynski}, \&
  {Piran}}]{Narayan_1992ApJ}
{Narayan}, R., {Paczynski}, B., \& {Piran}, T. 1992, \apjl, 395, L83

\bibitem[{{Nissanke} {et~al.}(2013){Nissanke}, {Holz}, {Dalal},
  {et~al.}}]{Nissanke2013arXiv}
{Nissanke}, S., {Holz}, D.~E., {Dalal}, N., {et~al.} 2013, astro-ph/1307.2638

\bibitem[{{Nuttall} \& {Sutton}(2010)}]{NuttallSutton2010PhRvD}
{Nuttall}, L.~K. \& {Sutton}, P.~J. 2010, \prd, 82

\bibitem[{{Nysewander} {et~al.}(2009){Nysewander}, {Fruchter}, \&
  {Pe'er}}]{Nysewander_2009}
{Nysewander}, M., {Fruchter}, A.~S., \& {Pe'er}, A. 2009, \apj, 701, 824

\bibitem[{{Ofek} {et~al.}(2006){Ofek}, {Kulkarni}, {Nakar},
  {et~al.}}]{Ofek2006ApJ}
{Ofek}, E.~O., {Kulkarni}, S.~R., {Nakar}, E., {et~al.} 2006, \apj, 652, 507

\bibitem[{{Ott}(2009)}]{Ott_ccSNreview_09}
{Ott}, C.~D. 2009, Class. Quant. Grav., 26, 063001

\bibitem[{{Paczynski}(1986)}]{Paczynski1986ApJ}
{Paczynski}, B. 1986, \apjl, 308, L43

\bibitem[{{Palaniswamy} {et~al.}(2014){Palaniswamy}, {Wayth}, {Trott},
  {McCallum}, {Tingay}, \& {Reynolds}}]{Palaniswamy2014ApJ}
{Palaniswamy}, D., {Wayth}, R.~B., {Trott}, C.~M., {McCallum}, J.~N., {Tingay},
  S.~J., \& {Reynolds}, C. 2014, \apj, 790, 63

\bibitem[{{Pannarale} \& {Ohme}(2014)}]{Pannarale2014ApJ}
{Pannarale}, F. \& {Ohme}, F. 2014, \apjl, 791, L7

\bibitem[{Pian {et~al.}(2006)}]{Pian_LLGRBs_06}
Pian, E. {et~al.} 2006, Nature, 442, 1011

\bibitem[{{Piran}(1999)}]{Piran_1999}
{Piran}, T. 1999, \physrep, 314, 575

\bibitem[{Piro \& Ott(2011)}]{Piro_Ott_2011}
Piro, A.~L. \& Ott, C.~D. 2011, The Astrophysical Journal, 736, 108

\bibitem[{Piro \& Pfahl(2007)}]{Piro_Pfahl_2007}
Piro, A.~L. \& Pfahl, E. 2007, The Astrophysical Journal, 658, 1173

\bibitem[{Piro \& Thrane(2012)}]{Piro_Thrane_2012}
Piro, A.~L. \& Thrane, E. 2012, The Astrophysical Journal, 761, 63

\bibitem[{{Predoi} {et~al.}(2012){Predoi}, {LIGO Scientific Collaboration},
  {Virgo Collaboration}, {Hurley}, \& {IPN Collaboration}}]{Predoi2012JPhCS}
{Predoi}, V., {LIGO Scientific Collaboration}, {Virgo Collaboration}, {Hurley},
  K., \& {IPN Collaboration}. 2012, Journal of Physics Conference Series, 363,
  012034

\bibitem[{Prestwich {et~al.}(2007)Prestwich, Kilgard, Crowther,
  {et~al.}}]{Prestwich_BH_IC10-X1_07}
Prestwich, A., Kilgard, R., Crowther, P., {et~al.} 2007, ApJ, 669, L21

\bibitem[{{Pshirkov} \& {Postnov}(2010)}]{Pshirkov2010}
{Pshirkov}, M.~S. \& {Postnov}, K.~A. 2010, \apss, 330, 13

\bibitem[{{Qin} {et~al.}(2013){Qin}, {Liang}, {Liang}, {et~al.}}]{Qin_2013}
{Qin}, Y., {Liang}, E.-W., {Liang}, Y.-F., {et~al.} 2013, \apj, 763, 15

\bibitem[{{Racusin} {et~al.}(2008){Racusin}, {Karpov}, {Sokolowski},
  {et~al.}}]{Racusin2008Natur}
{Racusin}, J.~L., {Karpov}, S.~V., {Sokolowski}, M., {et~al.} 2008, \nat, 455,
  183

\bibitem[{{Ravi} \& {Lasky}(2014)}]{Ravi2014MNRAS}
{Ravi}, V. \& {Lasky}, P.~D. 2014, \mnras, 441, 2433

\bibitem[{{Rees} \& {Meszaros}(1992)}]{Rees_1992}
{Rees}, M.~J. \& {Meszaros}, P. 1992, \mnras, 258, 41P

\bibitem[{{Rezzolla} {et~al.}(2011){Rezzolla}, {Giacomazzo}, {Baiotti},
  {Granot}, {Kouveliotou}, \& {Aloy}}]{Rezzolla_2011ApJ}
{Rezzolla}, L., {Giacomazzo}, B., {Baiotti}, L., {Granot}, J., {Kouveliotou},
  C., \& {Aloy}, M.~A. 2011, \apjl, 732, L6

\bibitem[{{Richards} {et~al.}(2011){Richards}, {Starr}, {Butler}, {Bloom},
  {Brewer}, {Crellin-Quick}, {Higgins}, {Kennedy}, \&
  {Rischard}}]{Richards2011}
{Richards}, J.~W., {Starr}, D.~L., {Butler}, N.~R., {Bloom}, J.~S., {Brewer},
  J.~M., {Crellin-Quick}, A., {Higgins}, J., {Kennedy}, R., \& {Rischard}, M.
  2011, \apj, 733, 10

\bibitem[{{Rosswog}(2005)}]{Rosswog2005ApJ}
{Rosswog}, S. 2005, \apj, 634, 1202

\bibitem[{{Rowlinson} {et~al.}(2013){Rowlinson}, {O'Brien}, {Metzger},
  {Tanvir}, \& {Levan}}]{Rowlinson2013MNRAS}
{Rowlinson}, A., {O'Brien}, P.~T., {Metzger}, B.~D., {Tanvir}, N.~R., \&
  {Levan}, A.~J. 2013, \mnras, 430, 1061

\bibitem[{{Rowlinson} {et~al.}(2010)}]{Rowlinson2010MNRAS}
{Rowlinson}, A. {et~al.} 2010, \mnras, 408, 383

\bibitem[{{Rowlinson} {et~al.}(2015)}]{Rowlinson2015}
{Rowlinson}, A. {et~al.} 2015, \mnras, in prep.

\bibitem[{{Sagiv} \& {Waxman}(2002)}]{Sagiv_2002}
{Sagiv}, A. \& {Waxman}, E. 2002, \apj, 574, 861

\bibitem[{{Sari} {et~al.}(1999){Sari}, {Piran}, \& {Halpern}}]{Sari_1999}
{Sari}, R., {Piran}, T., \& {Halpern}, J.~P. 1999, \apjl, 519, L17

\bibitem[{Sathyaprakash \& Schutz(2009)}]{SathyaprakashSchutz_LIVREV}
Sathyaprakash, B. \& Schutz, B.~F. 2009, Living Rev. Relativity, 12, 2

\bibitem[{Sathyaprakash(2004)}]{Saty04}
Sathyaprakash, B.~S. 2004, Invited review at ICRC2003, Tsukuba, Japan,
  preprint:gr-qc/0405136v1

\bibitem[{{Schmidt} {et~al.}(2005){Schmidt}, {Keller}, {Francis}, \&
  {Bessell}}]{Schmidt2005AAS}
{Schmidt}, B.~P., {Keller}, S.~C., {Francis}, P.~J., \& {Bessell}, M.~S. 2005,
  in Bulletin of the American Astronomical Society, Vol.~37, American
  Astronomical Society Meeting Abstracts \#206, 457

\bibitem[{Schutz \& Tinto(1987)}]{shultz_antenna_patt_fun_87}
Schutz, B. \& Tinto, M. 1987, MNRAS, 224, 131

\bibitem[{Schutz(1986)}]{shultz_chirp_86}
Schutz, B.~F. 1986, Nature, 323, 310

\bibitem[{{Schutz}(2011)}]{Schutz2011CQGra}
{Schutz}, B.~F. 2011, Classical and Quantum Gravity, 28, 125023

\bibitem[{{Shawhan}(2012)}]{Shawhan2012SPIE}
{Shawhan}, P.~S. 2012, in Society of Photo-Optical Instrumentation Engineers
  (SPIE) Conference Series, Vol. 8448, Society of Photo-Optical Instrumentation
  Engineers (SPIE) Conference Series

\bibitem[{Shibata \& Karino(2004)}]{shibataBarmode04}
Shibata, M. \& Karino, S. 2004, Phys. Rev. D, 70, 4022

\bibitem[{Singer(2015)}]{Singer_BAYSTAR_2015}
Singer, L. 2015, https://dcc.ligo.org/LIGO-P1500009

\bibitem[{{Singer} {et~al.}(2015){Singer}, {Kasliwal}, {Cenko},
  {et~al.}}]{SingerPTF2015ApJ}
{Singer}, L.~P., {Kasliwal}, M.~M., {Cenko}, S.~B., {et~al.} 2015, \apj, 806,
  52

\bibitem[{{Singer} {et~al.}(2014){Singer}, {Price}, {Farr},
  {et~al.}}]{Singer2014ApJ}
{Singer}, L.~P., {Price}, L.~R., {Farr}, B., {et~al.} 2014, \apj, 795, 105

\bibitem[{{Smith} {et~al.}(2014){Smith}, {Dekany}, {Bebek},
  {et~al.}}]{Smith_2014SPIE}
{Smith}, R.~M., {Dekany}, R.~G., {Bebek}, C., {et~al.} 2014, in Society of
  Photo-Optical Instrumentation Engineers (SPIE) Conference Series, Vol. 9147,
  Society of Photo-Optical Instrumentation Engineers (SPIE) Conference Series,
  79

\bibitem[{{Soderberg} {et~al.}(2006){Soderberg}, {Berger}, {Kasliwal},
  {et~al.}}]{Soderberg_2006}
{Soderberg}, A.~M., {Berger}, E., {Kasliwal}, M., {et~al.} 2006, \apj, 650, 261

\bibitem[{Soderberg {et~al.}(2006)}]{sodoburg_06_LLGRBRate_06}
Soderberg, A.~M. {et~al.} 2006, Nature, 422, 1014

\bibitem[{Somiya(2012)}]{kagra_2012}
Somiya, K. 2012, Classical and Quantum Gravity, 29, 124007

\bibitem[{Stanek(2003)}]{Stanek_LGRB_SN}
Stanek, K. 2003, \apj, 591, L17

\bibitem[{Steeghs \& Galloway(2015)}]{Comm_GOTO_2015}
Steeghs, D. T.~H. \& Galloway, D.~K. 2015, personal communication

\bibitem[{Sutton(2013)}]{sutton}
Sutton, P.~J. 2013, LIGO Document
  P1000041-v3,https://dcc.ligo.org/LIGO-P1000041

\bibitem[{{Sutton} {et~al.}(2010){Sutton}, {Jones}, {Chatterji},
  {et~al.}}]{Sutton_2010}
{Sutton}, P.~J., {Jones}, G., {Chatterji}, S., {et~al.} 2010, New Journal of
  Physics, 12

\bibitem[{{Tanaka} \& {Hotokezaka}(2013)}]{Tanaka2013ApJ}
{Tanaka}, M. \& {Hotokezaka}, K. 2013, \apj, 775, 113

\bibitem[{{Tanvir} {et~al.}(2013){Tanvir}, {Levan}, {Fruchter}, {Hjorth},
  {Hounsell}, {Wiersema}, \& {Tunnicliffe}}]{Tanvir2013Natur}
{Tanvir}, N.~R., {Levan}, A.~J., {Fruchter}, A.~S., {Hjorth}, J., {Hounsell},
  R.~A., {Wiersema}, K., \& {Tunnicliffe}, R.~L. 2013, \nat, 500, 547

\bibitem[{{Taylor} \& {Gair}(2012)}]{Taylor2012PhRvDb}
{Taylor}, S.~R. \& {Gair}, J.~R. 2012, \prd, 86, 023502

\bibitem[{{Taylor} {et~al.}(2012){Taylor}, {Gair}, \&
  {Mandel}}]{Taylor2012PhRvDa}
{Taylor}, S.~R., {Gair}, J.~R., \& {Mandel}, I. 2012, \prd, 85, 023535

\bibitem[{Thorne(1987)}]{thorn}
Thorne, K.~S. 1987, Gravitational Radiation, in 300 years of Gravitation,
  edited by S. Hawking and W. Israel, pages 330-458 (Cambridge, UK: Cambridge
  University Press)

\bibitem[{{Thornton} {et~al.}(2013)}]{Thornton2013Sci}
{Thornton}, D. {et~al.} 2013, Science, 341, 53

\bibitem[{Thrane \& Coughlin(2014)}]{Thrane_PhysRevD_2014}
Thrane, E. \& Coughlin, M. 2014, Phys. Rev. D, 89, 063012

\bibitem[{{Tingay} {et~al.}(2013)}]{Tingay2013PASA}
{Tingay}, S.~J. {et~al.} 2013, \pasa, 30, 7

\bibitem[{Tinney {et~al.}(2004)Tinney, Ryder, Ellis,
  {et~al.}}]{Tinney04iris2:a}
Tinney, C.~G., Ryder, S.~D., Ellis, S.~C., {et~al.} 2004, in Astronomical
  Telescopes and Instrumentation, SPIE Proc. 5492, 998--1009

\bibitem[{Tinto(1987)}]{tinto_antenna_patt_fun_87}
Tinto, M. 1987, MNRAS, 226, 829

\bibitem[{{Totani}(2013)}]{Totani2013PASJ}
{Totani}, T. 2013, \pasj, 65, L12

\bibitem[{{Troja} {et~al.}(2007){Troja}, {Cusumano}, {O'Brien},
  {et~al.}}]{Troja_2007}
{Troja}, E., {Cusumano}, G., {O'Brien}, P.~T., {et~al.} 2007, \apj, 665, 599

\bibitem[{{Tunnicliffe} {et~al.}(2014){Tunnicliffe}, {Levan}, {Tanvir},
  {et~al.}}]{Tunnicliffe2014MNRAS}
{Tunnicliffe}, R.~L., {Levan}, A.~J., {Tanvir}, N.~R., {et~al.} 2014, \mnras,
  437, 1495

\bibitem[{Urban(2015)}]{Comm_AlexUrban_2015}
Urban, A. 2015, personal communication

\bibitem[{Usov(1992)}]{Usov:1992zd}
Usov, V.~V. 1992, Nat, 357, 472

\bibitem[{{Usov} \& {Katz}(2000)}]{Usov_2000}
{Usov}, V.~V. \& {Katz}, J.~I. 2000, \aap, 364, 655

\bibitem[{{van Putten}(2008)}]{Putten2008ApJ}
{van Putten}, M.~H.~P.~M. 2008, \apjl, 684, L91

\bibitem[{{Veres} \& {M{\'e}sz{\'a}ros}(2014)}]{Veres_2014ApJ}
{Veres}, P. \& {M{\'e}sz{\'a}ros}, P. 2014, \apj, 787, 168

\bibitem[{{Vestrand} {et~al.}(2014){Vestrand}, {Wren}, {Panaitescu},
  {et~al.}}]{Vestrand2014Sci}
{Vestrand}, W.~T., {Wren}, J.~A., {Panaitescu}, A., {et~al.} 2014, Science,
  343, 38

\bibitem[{{Vestrand} {et~al.}(2005)}]{Vestrand2005Natur}
{Vestrand}, W.~T. {et~al.} 2005, \nat, 435, 178

\bibitem[{Virgilii {et~al.}(2008)Virgilii, Liang, \&
  Zhang}]{Virgilii_LLGRBs_08}
Virgilii, F.~J., Liang, E.-W., \& Zhang, B. 2008, MNRAS, 392, 91

\bibitem[{Was {et~al.}(2012)Was, Sutton, Jones, \& Leonor}]{SuttonPhysRevD2012}
Was, M., Sutton, P.~J., Jones, G., \& Leonor, I. 2012, Phys. Rev. D, 86, 022003

\bibitem[{{Wayth} {et~al.}(2015)}]{Wayth2015PASA}
{Wayth}, R.~B. {et~al.} 2015, \pasa, 32, 25

\bibitem[{Weisberg \& Taylor(1984)}]{Weisberg_Taylor_84}
Weisberg, J.~M. \& Taylor, J.~H. 1984, Phys. Rev. Lett., 52, 1348

\bibitem[{Wen \& Chen(2010)}]{wen10}
Wen, L. \& Chen, Y. 2010, Phys. Rev. D, 81, 082001

\bibitem[{Wen \& Schutz(2012)}]{advGWDet}
Wen, L. \& Schutz, B.~F. 2012, Chapter 5 in Advanced Gravitational Wave
  Detectors, ed. D.~G. Blair, E.~J. Howell, L.~Ju, \& C.~Zhao, Vol.~1
  (Cambridge University Press, Cambridge, UK)

\bibitem[{{Williams} {et~al.}(2012){Williams}, {Barthelmy}, {Denny}, {Graham},
  \& {Swinbank}}]{Williams2012SPIE}
{Williams}, R.~D., {Barthelmy}, S.~D., {Denny}, R.~B., {Graham}, M.~J., \&
  {Swinbank}, J. 2012, in Society of Photo-Optical Instrumentation Engineers
  (SPIE) Conference Series, Vol. 8448, Society of Photo-Optical Instrumentation
  Engineers (SPIE) Conference Series, 0

\bibitem[{{Wilson} {et~al.}(2011){Wilson}, {Ferris}, {Axtens},
  {et~al.}}]{Wilson2011}
{Wilson}, W.~E., {Ferris}, R.~H., {Axtens}, P., {et~al.} 2011, \mnras, 416, 832

\bibitem[{Woosley \& Janka(2005)}]{WoosleyJankaNature_05}
Woosley, S. \& Janka, T. 2005, Nature Physics, 1, 147

\bibitem[{{Woosley} \& {Bloom}(2006)}]{WB_06}
{Woosley}, S.~E. \& {Bloom}, J.~S. 2006, Annu. Rev. Astro. Astrophys., 44, 507

\bibitem[{{Woosley} {et~al.}(1999){Woosley}, {MacFadyen}, \&
  {Heger}}]{Woosley_99}
{Woosley}, S.~E., {MacFadyen}, A.~I., \& {Heger}, A. 1999, Supernovae and
  Gamma-Ray Bursts (Cambridge, UK: Cambridge University Press)

\bibitem[{{Zhang}(2013)}]{Zhang2013ApJ}
{Zhang}, B. 2013, \apjl, 763, L22

\bibitem[{{Zhang}(2014)}]{Zhang2014ApJ}
{Zhang}, B. 2014, \apjl, 780, L21

\bibitem[{{Zhang} \& {M{\'e}sz{\'a}ros}(2001)}]{Zhang2001ApJ}
{Zhang}, B. \& {M{\'e}sz{\'a}ros}, P. 2001, \apjl, 552, L35

\bibitem[{{Zink} {et~al.}(2012){Zink}, {Lasky}, \& {Kokkotas}}]{Zink2012PhRvD}
{Zink}, B., {Lasky}, P.~D., \& {Kokkotas}, K.~D. 2012, \prd, 85, 024030

\end{thebibliography}
\end{document}